\documentclass[12pt]{iopart}

\usepackage{graphicx}
\usepackage{caption}
\usepackage{natbib}
\usepackage{indentfirst}
\usepackage[label font=bf,labelformat=simple]{subfig}
\usepackage{floatrow}

\usepackage{hyperref}

\setcitestyle{square}
\setcitestyle{numbers}
\setcitestyle{comma}
\begin{document}

\title[]{Predictive study of non-axisymmetric neutral beam ion loss on the upgraded KSTAR plasma-facing components} 
\author{Taeuk Moon$^1$, Tongnyeol Rhee$^2$*, Jae-Min Kwon$^2$*, Young-Mu Jeon$^2$, and Eisung Yoon$^1$*}
\address{$^1$ Ulsan National Institute of Science and Technology, Ulsan 44919, Republic of Korea}
\address{$^2$ Korea Institute of Fusion Energy, Daejeon 34133, Republic of Korea}

\ead{\mailto{trhee@kfe.re.kr}, \mailto{jmkwon74@kfe.re.kr}, \mailto{esyoon@unist.ac.kr}}
\vspace{10pt}
\begin{indented}
\item[] 
\end{indented}

\begin{abstract}

We simulate ion loss induced by neutral beam injection (NBI) in three-dimensional (3D) space with high fidelity on the plasma-facing components (PFCs) of the Korea Superconducting Tokamak Advanced Research (KSTAR) device.
Utilizing a 3D collision detection routine added to the NuBDeC code and computer-aided design data reflecting the recent upgrade to a tungsten divertor, we have characterized 3D heat flux distribution and patterns over PFCs due to the NB ion loss throughout parameter scans. First, we identify axial asymmetry in the heat flux distribution.
The plasma-wetted areas extend along the direction of the plasma current and deviate diagonally in the direction of the NB ion's poloidal turn.
Additionally, on the PFC surfaces, local heat flux peaks are observed along both poloidal and toroidal directions.
These local peaks emerge on the surfaces of the PFCs that protrude toward the regions swept by the NB ions.
Second, through a case study of NB1-C (beam source C of neutral beam 1) that results in the most loss, we analyze several changes in the heat flux patterns observed on the divertor and the poloidal limiter.
Such analysis allows us to examine how variations in the parameters affect the movement of the peak heat flux position and the extent of the plasma-wetted area, if formed.
We observe that the ion loss increases under several conditions: shallower beam deposition, higher beam energy, larger poloidal beta, and lower plasma current.
Ions born near the plasma edge due to shallow beam injection follow orbits with a larger minor radius.
Flux surface shifts and ion drifts from changes in beam energy, poloidal beta, and plasma current move these orbits closer to the wall. 
These factors increase the chance of ion loss through wall collisions.
This study is believed to help optimize design and operation of NBI systems.
\end{abstract}

%
%
%
%
%

\section{Introduction}

The Korea Superconducting Tokamak Advanced Research (KSTAR) device is dedicated to investigating steady-state high-$\beta$ plasma operations and advanced tokamak physics relevant to fusion. 
As a part of these investigations, various advanced scenarios have been examined for H-mode operations.
During these scenarios, neutral beam injection (NBI), electron cyclotron heating, and ion cyclotron heating systems have been utilized for the selective heating of ions and electrons \cite{kim2023development, han2023kstar, kwak2007progress}.

However, in typical H-mode scenarios, NBI heating can lead to an accumulation of heat load on the plasma-facing components (PFCs), posing the risk of overheating and potential damage to the device. 
Various studies have aimed to understand the causes of this overheating through NBI heating simulations \cite{kurki2016protecting, kurki2008fast, ward2022locust, kurki2018clearing, akers2018high}.
In particular, research on the KSTAR device has investigated the effects of NB ion loss using both numerical methods and experiments \cite{Rhee_2019, rhee2022simulation, kim2018enhanced}. 
One of the reasons for the overheating was identified as the NB ion loss \cite{Rhee_2019, kim2016prompt}.

Previous simulations have used simplified wall geometry \cite{kim2016prompt, bak2017measurement, oh2016overview}, and further research could explore the patterns of non-axisymmetric NB ion loss over the non-axisymmetric geometry of real-world fusion device PFCs.
To better evaluate the loss across the diverse features of such components, it is beneficial to incorporate actual geometry from three-dimensional (3D) computer-aided design (CAD) data into the calculations.
In that regard, neutral beam deposition and orbit-following simulation code (NuBDeC) \cite{Rhee_2019} has been equipped the realistic wall geometry with collision detection capability in 3D space \cite{moon4537156development}.

Meanwhile, the lower carbon divertor of KSTAR has recently been upgraded to a tungsten divertor to handle the high peak heat load \cite{hong2020brief, kwon2021cfd}.
In this case, NB ions possessing high energy can potentially cause substantial sputtering of tungsten atoms from the new divertor surface, which leads to material erosion, thereby shortening the lifetime of the divertor. Moreover, it can introduce tungsten impurities into the plasma, which may result in severe energy loss by Bremsstrahlung radiation.
Capturing where heat accumulates on the new tungsten divertor can help analyze origin of these interactions. 
Also, it can provide several means to ensure the durability of the PFCs and to maintain optimal plasma conditions for typical H-mode scenarios.

This research models NB ion loss for the KSTAR PFCs including the newly upgraded tungsten divertor using NuBDeC.
In the results, various non-axisymmetric loss patterns are identified on PFCs that consistently emerge across a range of beam energies, beam sources, and magnetic equilibria during a parameter scan for the ion loss simulation.

In Section~\ref{sec:ModelingAndSimulationSetups}, we describe the experimental process and present an overview of the algorithms used to detect NB ion loss on the 3D geometry of the PFCs. 
To speed up the simulation, we introduce a new beam deposition method with selective beam particle sampling. 
Also, the finite Larmor radius effect is examined for a more realistic calculation by comparing the results of the simulations between Lorentz orbit and the guiding-center-following orbit.
We then briefly describe how to calculate the heat flux distribution of NB ion loss.
The simulation setup parameters are outlined, namely the plasma profile, magnetic equilibrium, and other tokamak plasma parameters along with their ranges for the parameter scan to analyze changes in the heat flux distribution.

In Section~\ref{sec:Results}, we explore the heat flux distribution of NB ion loss calculated by the NuBDeC simulation. 
We show 3D heat flux distribution over the divertor and poloidal limiter (PL) unfolded on two-dimensional (2D) planes. 
We also investigate the non-axisymmetric patterns observed in the reference setup and track the variations in heat flux distribution, focusing on the effects of particle drift motion and the non-axisymmetric shape of the tungsten divertor.

Finally, Section~\ref{sec:Conclusion} concludes with a summary of the findings and potential areas for future research.

\section{Modeling and Simulation Setups} \label{sec:ModelingAndSimulationSetups}
The overall experimental process comprises following simulation and post-processing steps:
\begin{enumerate}
        \item NB ionization, 
        \item Orbit following and collision detection with the 3D wall, and
        \item Calculation of heat flux distribution and its analysis.
\end{enumerate}
The first two steps correspond to the sequence of modeling the lifetime of wall-colliding NB ions.
Subsequently, heat flux due to NB ion loss is calculated based on the collision-to-wall computations.
Then, the maximum heat flux value and the power deposition for each categorized PFC regions are analyzed.
The upcoming subsections illustrate details of each procedure and the simulation configurations employed for the parameter scan.

\subsection{Selective beam deposition} \label{subsec:beam_selection}
To enhance the computational efficiency of NB ion loss simulations, it is advantageous to avoid tracking well-confined NB ion orbits among the beam particles, whose ionization locations and velocities are determined as described in Reference~\cite{Rhee_2019}.
This selection process is executed in the 2D phase space spanned by the conserved quantities of toroidal canonical momentum, $P_{\phi}=mv_{||}g/B-q\psi_{p}$, and magnetic moment, $\mu=\frac{1}{2B}mv_{\bot}^{2}$, for an axisymmetric toroidal system \cite{White_book}.
Here, $m$ and $q$ denote the particle's mass and charge, $v_{\parallel}$ and $v_{\perp}$ represent speeds parallel and perpendicular to the  magnetic field $B$, and $\psi_p$ is the poloidal magnetic flux.
In cylindrical coordinates, the poloidal current function is given by $g = RB_{\phi}$, where $R$ is the radial coordinate from the major axis of the tokamak and $B_{\phi}$ is the toroidal magnetic field.
We note that each particle's 5D guiding-center-following orbit for a given energy is represented as a point in this phase space.
This enables us to determine outermost radius of particles within the tokamak with minimal computational effort. 

In NuBDeC, ionized particles undergo a test of their orbits' radial positions at the midplane on the LFS.
To filter out particles sufficiently distant from the wall, we defined a critical radial position as the intersection of the user-selected magnetic flux surface and the outboard midplane.
If the outermost radial position calculated for a testing ion is closer to the magnetic axis than the user-defined critical radial position, the NB ion is safely considered as well-confined and excluded from a list of particles for orbit tracing. Typically, the co-passing particles reach the outermost radial position at the outboard midplane.

\begin{figure}\centering
        \sidesubfloat[]{\includegraphics[height = 2.5in]{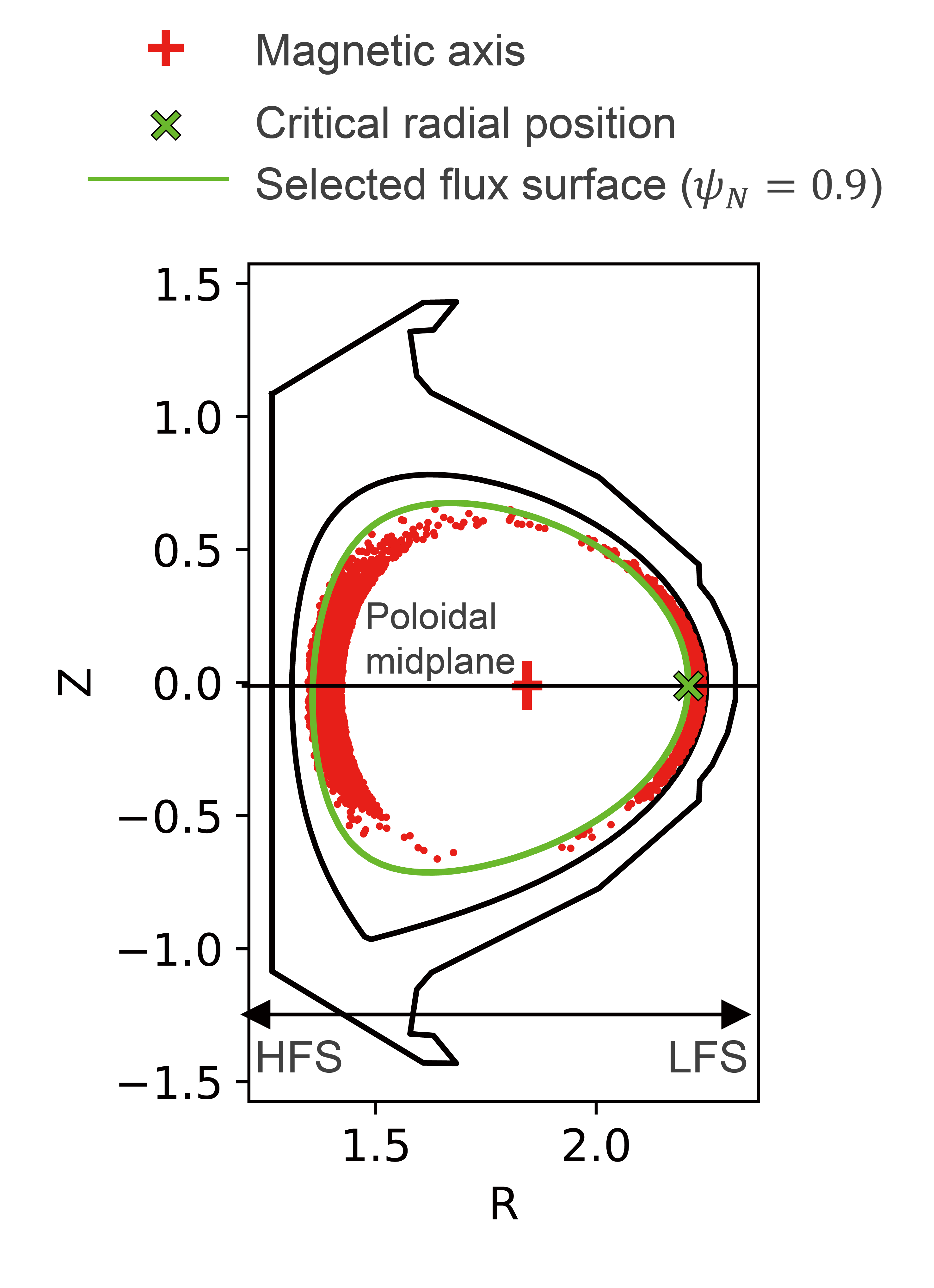}\label{fig:SelectionA}}
        \sidesubfloat[]{\includegraphics[height = 2.5in]{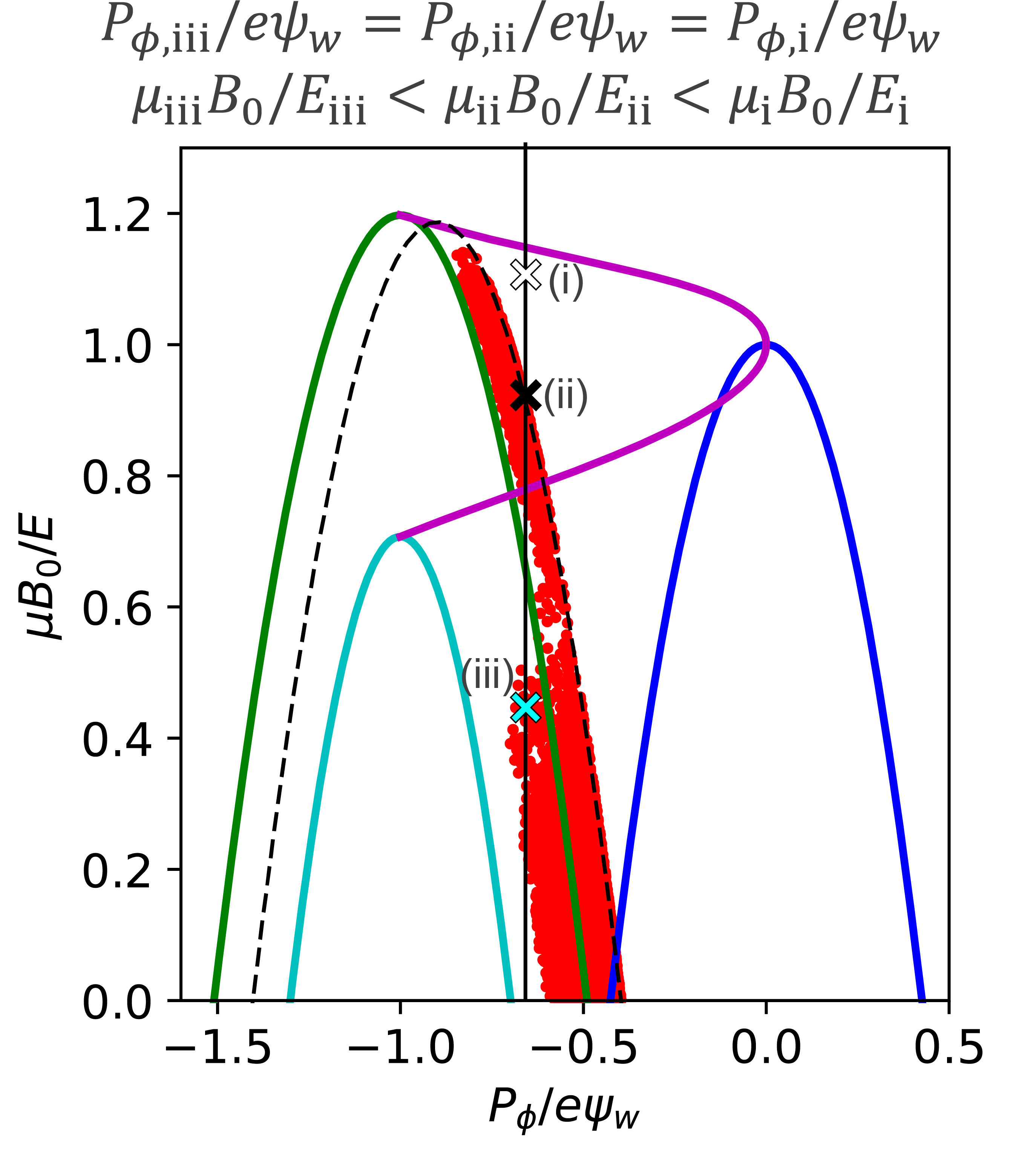}\label{fig:SelectionB}}
        \caption{(a) Selected beam deposition positions (marked by red dots) in the toroidal cross section.
        The flux surface (light green closed curve, $\psi_N = 0.9$) intersects the poloidal midplane (black horizontal line) at the critical radial position on the LFS (light green x-mark).
        (b) The initial states of individual NB ions included in the beam sampling at the moment of their ionization shown in the phase space. 
        In the phase space, magenta, blue, green, cyan, and black dashed lines respectively represent the phase space points that are capable of passing the trapped boundary, magnetic axis, LCFS at the LFS midplane, LCFS at the HFS midplane, and $\psi_N=0.9$ at the LFS midplane. $E$ and $e$ are NB ion energy and electron charge.
        }
        \label{fig:Selection}
\end{figure}

Figure~\ref{fig:Selection} shows the $P_\phi - \mu$ diagram for $\psi_N = 0.9$ with the critical radial position. 
For this example, the simulation is set with a 100 keV NB1-C using the reference magnetic equilibrium (see Section~\ref{subsec:LorentzOrbits}). 
The red dots in the phase space of Figure~\ref{fig:SelectionB} indicate the initial states of particles whose paths cross the poloidal midplane at or beyond the critical radial position on the LFS side. 
These particles are considered to be not well-confined and to be included in particle tracing due to their potential loss. 
It is noticeable that some of the selected particles are inside the flux surface in the high-field side (HFS). 
This is because the particles drift out across the flux surface to reach beyond the critical radial position on the LFS of the poloidal midplane.

A particle ionized at $\psi_p$, $P_{\phi,p}$, $\mu_p$, and $E_p$ is accepted from this beam sampling for orbit following stage if its $\mu_p B_0/E_p$ value lies below the quadratic line defined by Equation~\ref{eq:Phase_space_equation} for the user-defined magnetic flux $\psi_c$, denoted as a black dashed line in Figure~\ref{fig:SelectionB}, when considering the particle .
In other words, if a particle is assumed to be at the critical radial position with \(P_{\phi,p}\), and its \(\mu_p B_0/E_p\) value is smaller than the \(\mu_c B_0/E_p\) value that satisfies Equation~\ref{eq:Phase_space_equation} for the magnetic field \(B_c\) and poloidal current function \(g_c = g(\psi_c)\) at that assumed position, then the particle is used for beam tracing.
Here, $E_p$ and $B_0$ are the particle's energy and the magnetic field strength at the magnetic axis, respectively.

In Figure~\ref{fig:SelectionB}, the initial states of three particles with the same $P_\phi/e\psi_w$ value are compared in the phase space. 
$\psi_w$ is the $\psi_p$ difference between magnetic field flux at last closed flux surface (LCFS) and magnetic axis.
Point (ii) represents the initial state that satisfies conservation of energy and canonical toroidal angular momentum for $\psi_N = 0.9$, i.e., 
\begin{equation}\label{eq:Phase_space_equation} 
        \mu_c B_c = E-\frac{\left(P_{\phi c}+q\psi_{c}\right)^{2}B_c^{2}}{2mg_c^{2}},
\end{equation}
where $\psi_{N,p} = (\psi_p - \psi_{\textrm{magnetic axis}})/(\psi_{\textrm{LCFS}} - \psi_{\textrm{magnetic axis}})$ is normalized poloidal magnetic flux.
Comparing $\mu_{\textrm{(iii)}} B_0/E_{\textrm{(iii)}}$ with $\mu_{\textrm{(ii)}} B_0/E_{\textrm{(ii)}}$, we find that $\mu_{\textrm{(iii)}} B_0/E_{\textrm{(iii)}}$ is relatively smaller.
Particles with such initial motion states are included in beam tracing.

Well-confined NB ions is not used for beam sampling, but the number of them is stored for calculating the sample weight, or the physical particle number per sample particle.
In NuBDeC simulation, users select the desired beam source and specify the power and beam energy for each setup. This determines the actual number of NB particles that need to be injected by each beam source. 
Based on the total number of physical particles, the weight is calculated as the portion of the total physical particles to show the importance of each selected or ignored sample particle in terms of its energy contribution to the overall heat flux distribution.

\subsection{Collision detection}

\begin{figure}[H]\centering
\includegraphics[width=0.9\textwidth]{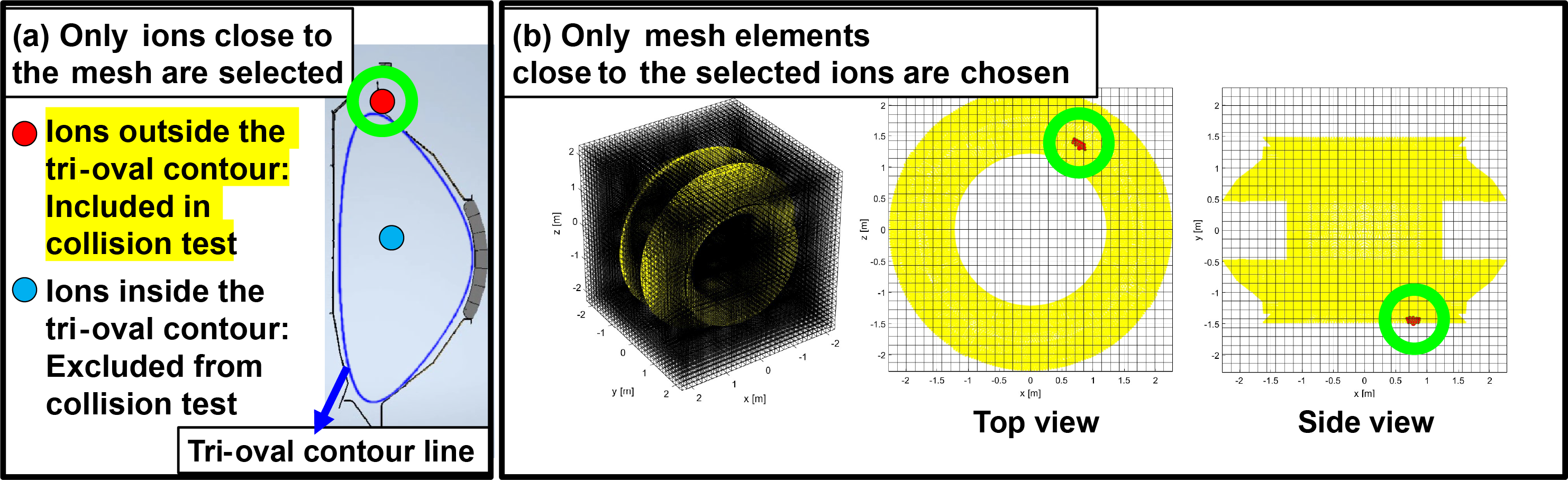}
        \captionsetup{width=\linewidth, format=hang}
        \caption{
                Broad phase of the collision detection steps. (a) The blue tri-oval divides the collision detection domain into two parts: inside and outside the tri-oval. The red dot in the green circle is included in the collision test queue, selecting ions near the wall boundary. (b) The mesh domain is divided into a uniform grid drawn over the yellow mesh. The small red area in the green circle marks the mesh elements in the same grid cells as the ions selected in (a). Collision detection tests are performed only between the ions selected in (a) and the mesh elements in (b).
                }
        \label{fig:CollisionDetectionProcessBroad}
\end{figure}

\begin{figure}[H]\centering
\includegraphics[width=0.9\textwidth]{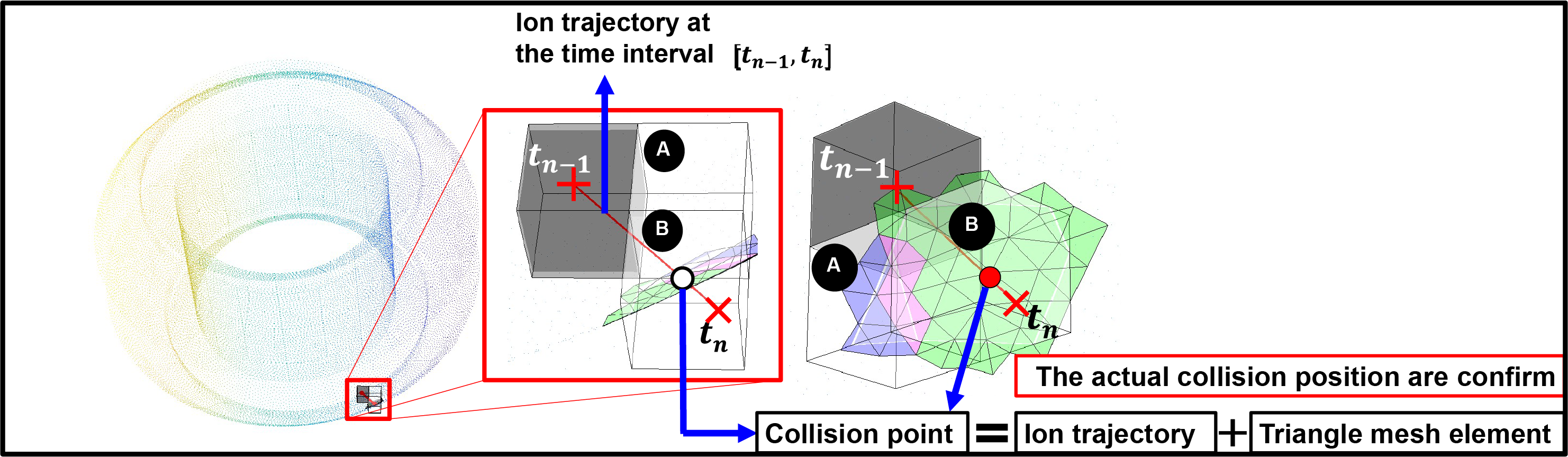}
        \captionsetup{width=\linewidth, format=hang}
        \caption{
                Narrow phase of the collision detection steps. In the first step of the previous broad phase, three uniform grid cells in the red square were selected from the overall mesh [Figure~\ref{fig:CollisionDetectionProcessBroad}(b)]. The ion trajectory segment (red line from the plus sign to the $\times$ symbol during the time step [$t_{n-1}$, $t_n$] of NuBDeC simulation) passes through these three cells. In the narrow phase, using the ray-casting method, collisions with nearby mesh elements are checked along the segment. The white circle indicates the location of a collision. The blue and green triangles are located in grid cells A and B, respectively, while the purple triangles are in both cells.
                }
        \label{fig:CollisionDetectionProcessNarrow}
\end{figure}

In each NuBDeC simulation time step, NB ion loss is captured by detecting collisions between the KSTAR PFC surfaces and a path segment, which is formed by drawing a straight line between the previous and current positions of the NB ion. 
This study employs an unstructured mesh to accurately represent the CAD geometry of KSTAR PFC surfaces. Unlike structured meshes, which comprise regularly arranged cells (typically quadrilaterals in 2D), unstructured meshes offer superior flexibility in capturing the intricate geometry like KSTAR PFC components. To reduce the number of pairwise collision detections, efficient algorithms for unstructured meshes are implemented in the NuBDeC code \cite{moon4537156development}.
Figure~\ref{fig:CollisionDetectionProcessBroad} and \ref{fig:CollisionDetectionProcessNarrow} show the collision detection algorithm utilized in this study for the broad and narrow phases, respectively \cite{ericson2004real}. The complete collision detection process for NuBDeC including the algorithms used in each phase is detailed in Reference \cite{moon4537156development}. 

For fast collision detection, in this work, we adopt a tri-oval contour and a uniform grid in the broad phase to reduce area subject to collision test and consequently sort out ion particles only that has finite possibility to collide with PFCs.
To quickly retrieve the list of mesh elements in each chosen grid cell, we use a hash table that maps each cell of the fictitious uniform grid to unstructured mesh elements touching the corresponding cell. In the narrow phase, we use the ray-casting method to find an intersecting point between an ion path segment and a PFC unstructured mesh element.

Figure~\ref{fig:CollisionDetectionProcessBroad} shows the broad phase, where we select ion path segments near the mesh while excluding those within the tri-oval contour (far from the mesh) and pick the mesh elements near the remaining ion path segments. 
Figure~\ref{fig:CollisionDetectionProcessNarrow} describes how the ray-casting method identifies collision points between the selected ion path segments and the nearby mesh elements during each NuBDeC simulation time step between [$t_{n-1}$, $t_n$].


\begin{figure}[H]\centering
\includegraphics[width=0.7\textwidth]{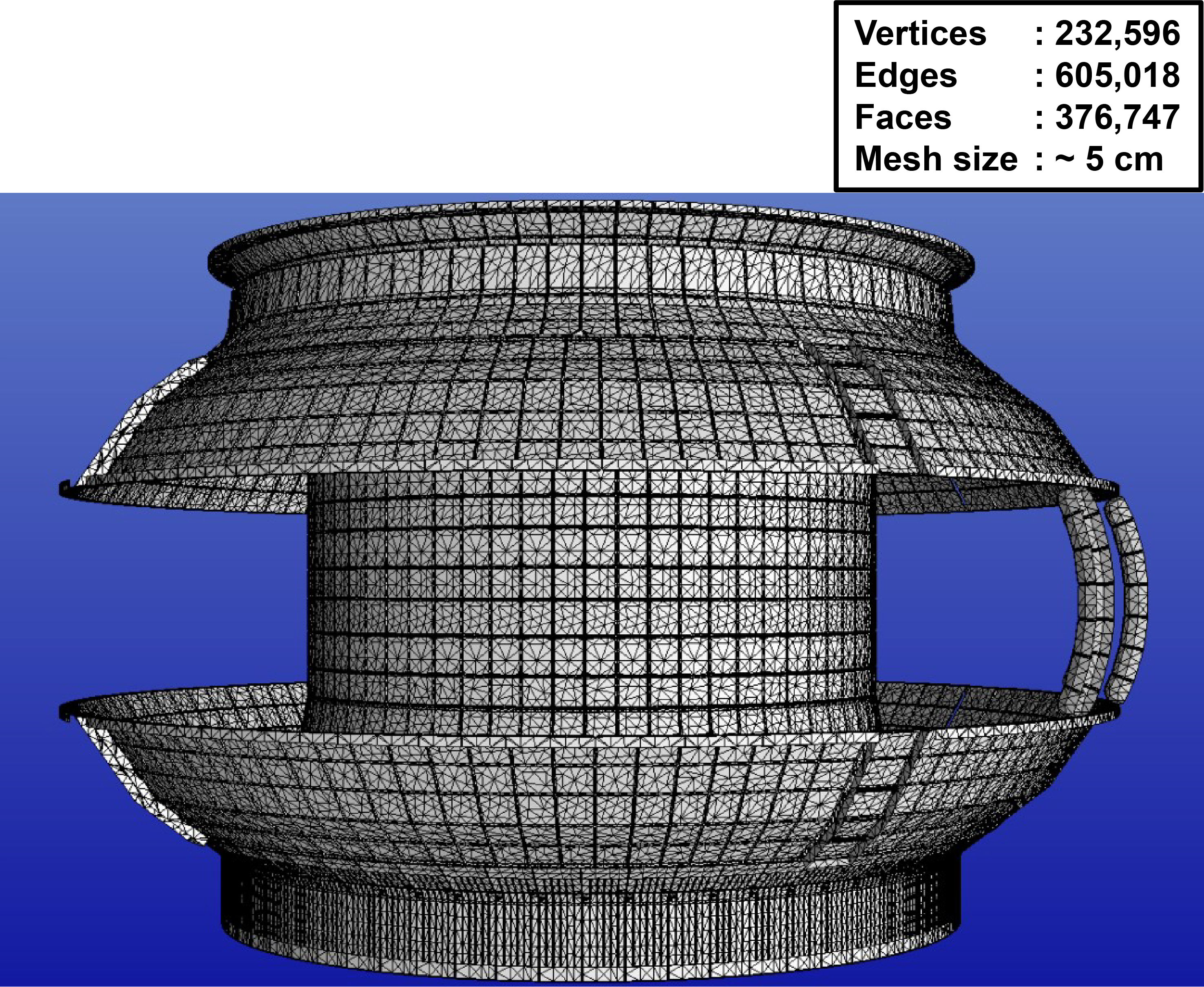}
        \captionsetup{width=\linewidth, format=hang}
        \caption{
                A mesh for the KSTAR PFC geometry used in the 3D collision detection to calculate the positions of NB ion loss.
                }
        \label{fig:KSTARmesh}
\end{figure}

The unstructured mesh used for the collision detection is built by triangular tessellation (376,747 triangles, each of size $\sim$5 cm) over the defeatured PFC shape.
In the tessellation on the geometry of the KSTAR device, the mesh is built only on the surfaces of the PFCs and their leading edges that NB ions are likely to collide with. 
Figure~\ref{fig:KSTARmesh} shows the NuBDeC mesh for this study obtained by these procedures.

\subsection{NB ions following Lorentz orbits} \label{subsec:LorentzOrbits}
\begin{figure}[H]\centering
\includegraphics[width=0.7\textwidth]{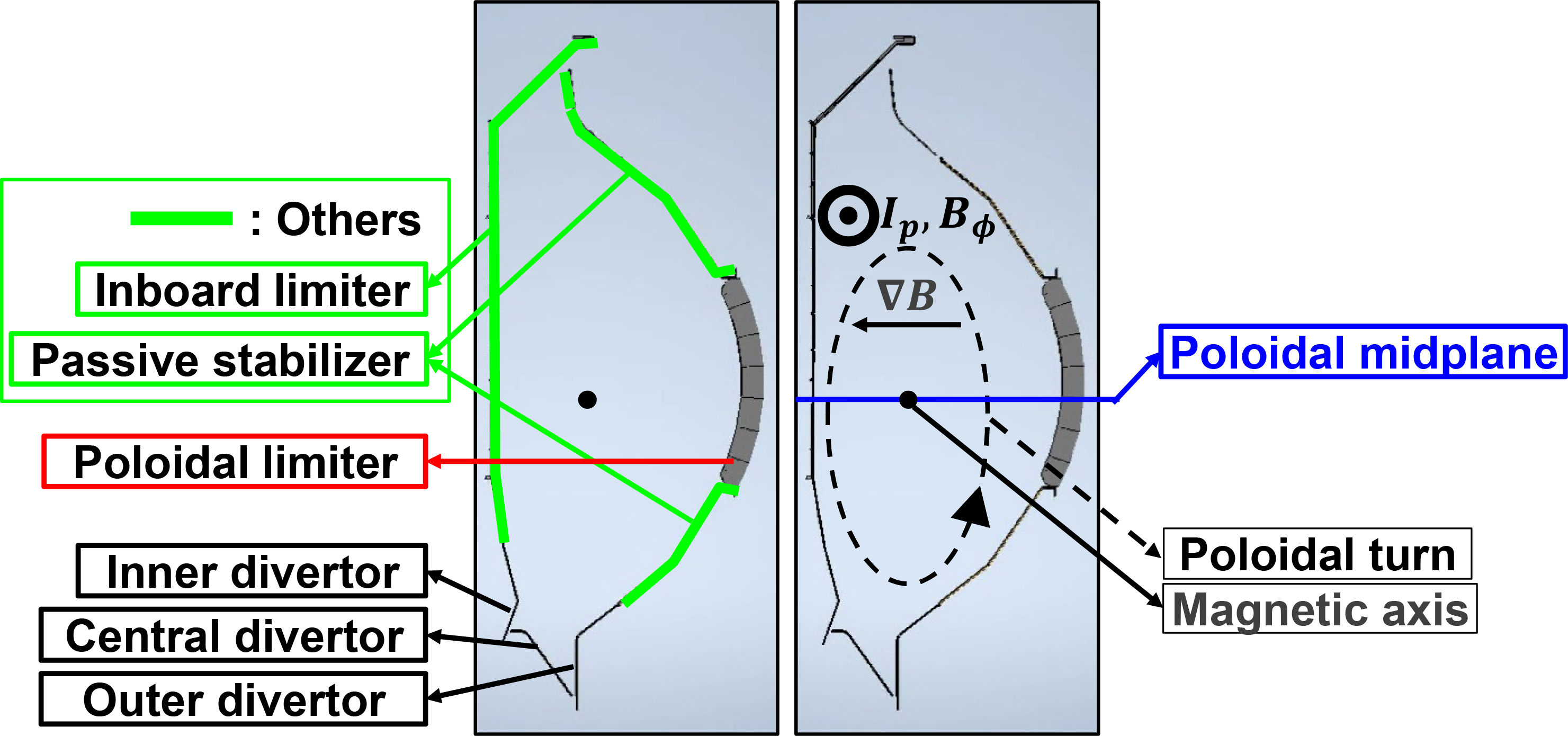}
        \captionsetup{width=\linewidth, format=hang}
        \caption{
                 Toroidal cross section of KSTAR PFC surfaces illustrating the poloidal turn, poloidal midplane, and the following PFC regions, used for categorizing the simulation results: divertor, PL, and others, which include the plasma-facing surface of passive stabilizer and inboard limiter.
                }
        \label{fig:poloidalCrossSectionView}
\end{figure}

The orbit of a NB ion projected on the toroidal cross section (or poloidal plane) of KSTAR PFCs is shown in Figure~\ref{fig:poloidalCrossSectionView}. 
The NB ions born from ionization of neutral particles turn poloidally at different times.
When particles collide with the wall, the collision is recorded as a loss in each of the PFC regions categorized as divertor, PL, and others as described in Figure~\ref{fig:poloidalCrossSectionView} (see Section~\ref{subsec:parameterScan} for more details).
For NB ions with high energy, the Larmor radius is expected to be large due to their high energy, indicating an expanded area traversed by the particles' gyration motion, extending from the guiding center to the Larmor radius. 
This gyration motion increases the likelihood of collisions with the 3D walls. 
To account for the contribution of gyration, the NB ions are simulated according to the Lorentz orbit: 
\begin{equation}\label{eq:LorentzdXdt}
\eqalign{\frac{d\vec{x}}{dt} &= \vec{v},  \\
\frac{d\vec{v}}{dt} &= \frac{q}{m} \left(\vec{v} \times \vec{B} \right), } 
\end{equation}
where $q$ and $m$ represent the NB ion charge and mass respectively, and $\vec{B}$ denote the magnetic field. 
The finite Larmor radius effect of NB ions can be evaluated by comparing the simulation results of the full (Lorentz) orbit with those of the guiding-center orbit \cite{Rhee_2019}.

\begin{figure}[H]\centering
    \sidesubfloat[]{\includegraphics[width=0.7\textwidth]{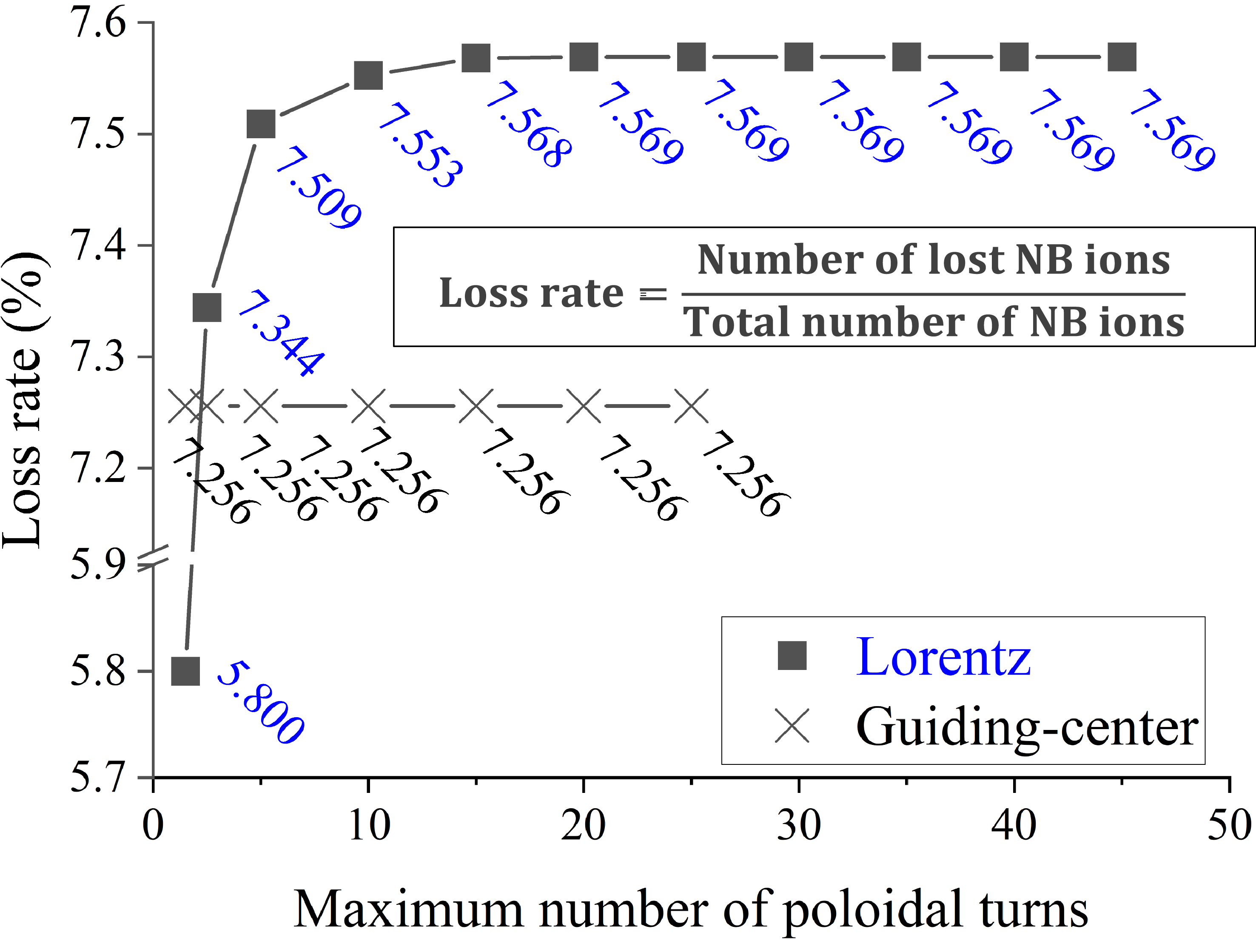}\label{fig:convergenceTestFigureA}}
    \hfil
    \sidesubfloat[]{\includegraphics[width=0.7\textwidth]{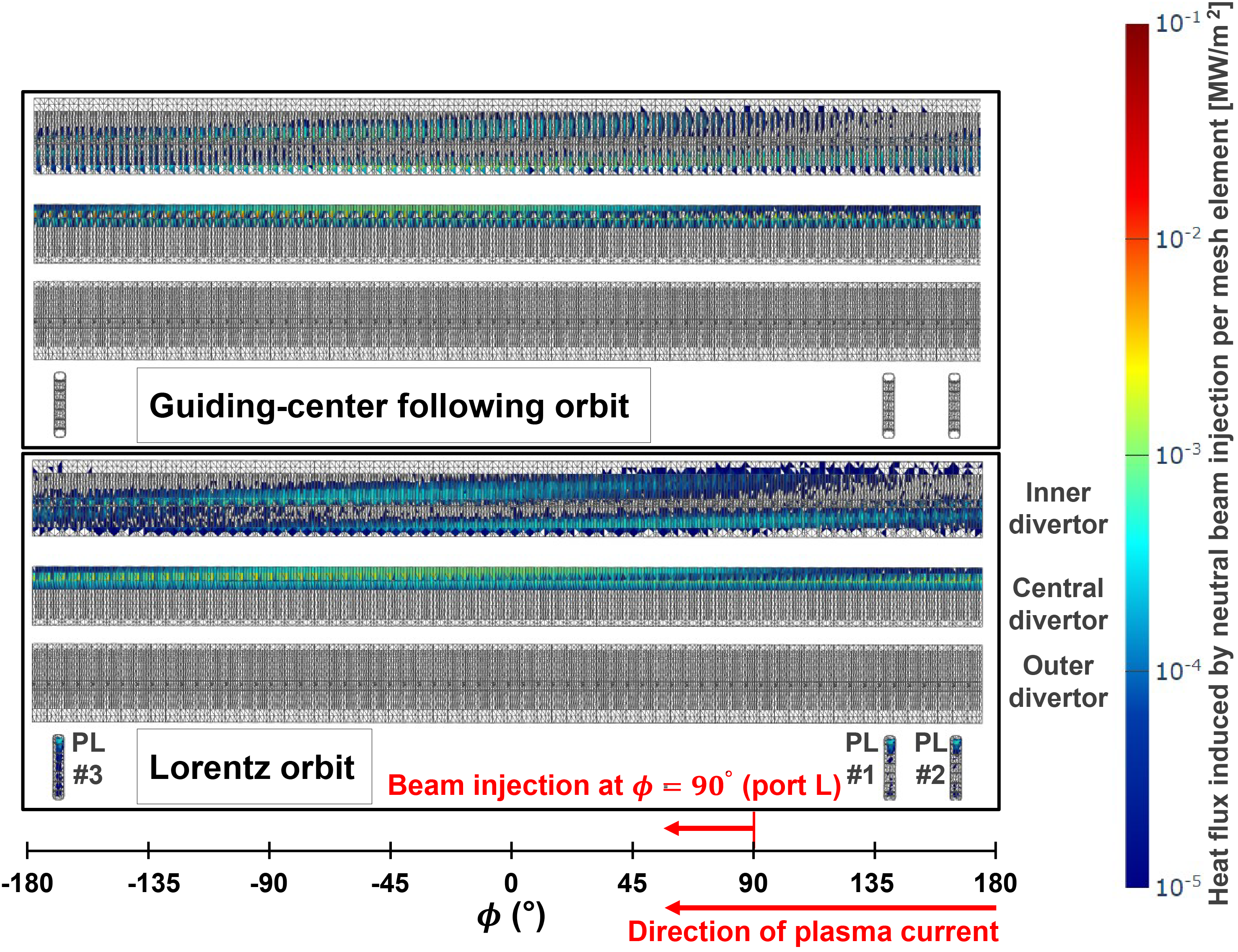}\label{fig:convergenceTestFigureB}}
    \caption{
            (a) Convergence test for the loss rate of NB ions on the KSTAR PFCs across NuBDeC simulations with reference setup described in Section~\ref{sec:Results} for increasing numbers of allowed poloidal turns. (b) Comparison of heat flux distributions due to NB ion losses for the guiding center and Lorentz orbits, respectively.
            }
    \label{fig:convergenceTestFigure}
\end{figure}

We compare the Lorentz and guiding-center-following orbits for convergence of loss rate in the reference setup described in Section~\ref{sec:Results}.
First, the loss rate for the ions following Lorentz orbits is found to be converging at approximately $20$ poloidal turns, as shown in Figure~\ref{fig:convergenceTestFigureA}.
On the other hand, the orbit following the guiding-center converges before $5$ poloidal turns.
In addition, the Lorentz orbit case results in 0.3\% more particle loss than that of the guiding-center-following orbit.
However, the heat flux distributions of the two cases significantly differ as shown in Figure~\ref{fig:convergenceTestFigureB}.
Thus, in this study we employ the full-orbit path tracing for all parameter scans for NB ion loss estimation.


\begin{figure}[H]\centering
    \includegraphics[width=0.9\textwidth]{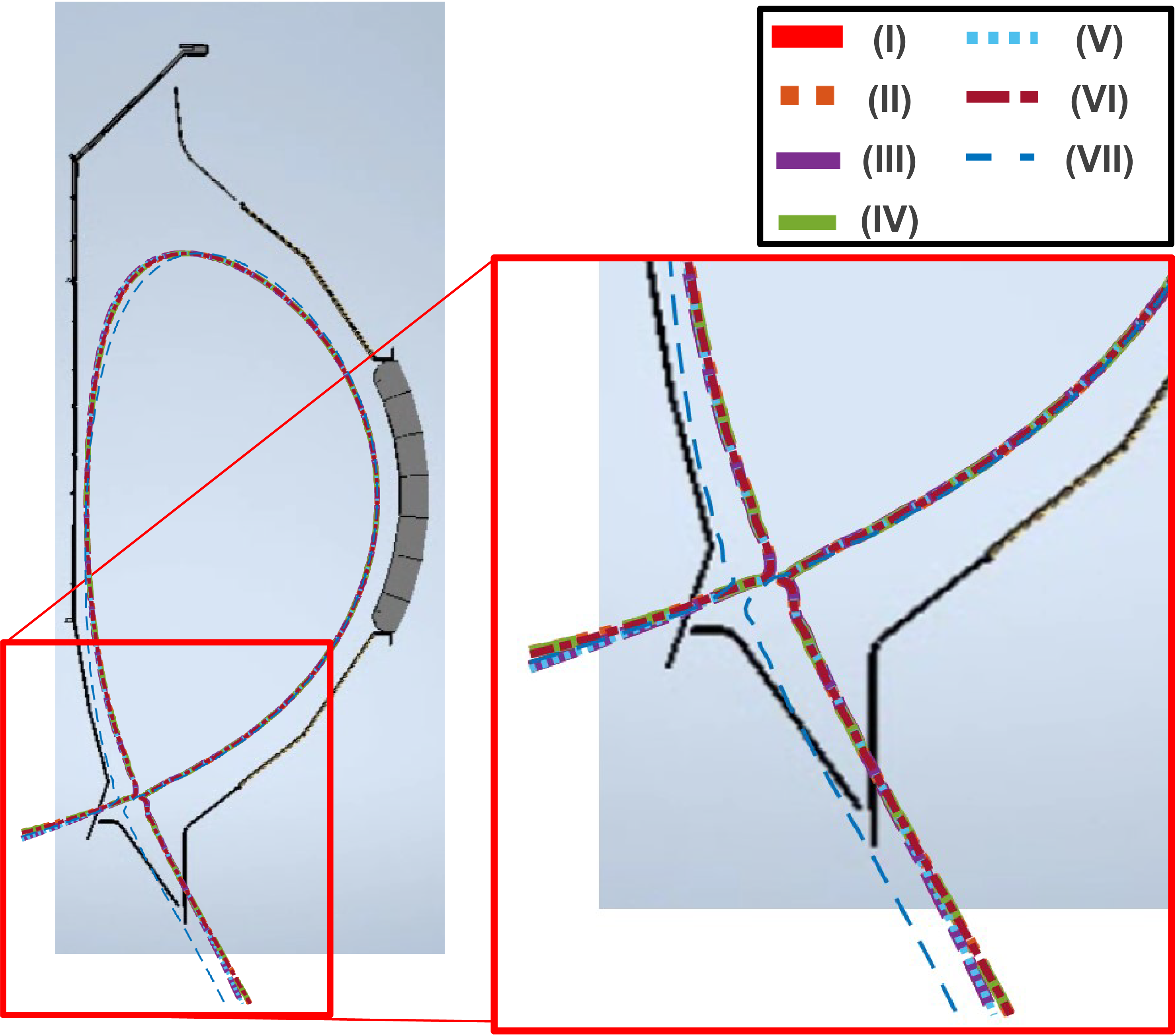}
    \caption{
            Toroidal cross section of magnetic separatrices for seven magnetic equilibria (Table~\ref{table:magneticFieldtable}) in the NuBDeC simulation parameter scan. Equilibrium case (I) is the reference setup.
            }
    \label{fig:magneticFieldfigure}
\end{figure}

\begin{table}[H]
\caption{Plasma current ($I_p$), poloidal beta ($\beta_p$), and the strike point position of the outer separatrix leg for each magnetic equilibrium illustrated in Figure~\ref{fig:magneticFieldfigure}.}
\label{table:magneticFieldtable}
\begin{indented}
\lineup
\item[]\begin{tabular}{@{}*{4}{c}}
\br
Equilibrium case & $I_p$ & $\beta_p$ & strike point position\\
\mr
(I)        & 1.00~MA & 1.0 & Outer divertor \\
(II)        & 1.00~MA & 2.0 & Outer divertor \\
(III)        & 0.75~MA & 1.0 & Outer divertor \\
(IV)        & 0.75~MA & 2.0 & Outer divertor \\
(V)        & 0.50~MA & 1.0 & Outer divertor \\
(VI)        & 0.50~MA & 2.0 & Outer divertor \\
(VII)        & 1.00~MA & 1.0 & Central divertor \\
\br
\end{tabular}
\end{indented}
\end{table}

\subsection{Post-processing}\label{sec:postProcesssing}
After the simulation, we calculate the heat flux on the $k^{th}$ mesh element ($\Gamma_k$) by aggregating the heat flux transferred by the $i^{th}$ ion to the $k^{th}$ mesh element ($\Gamma_{ik}$) as follows.

\begin{equation*}
\eqalign{E_i &= \frac{1}{2}\omega_i m_i \| \vec{v_i} \|^2,  \\
\Gamma_{ik} &= \frac{E_i}{A_k\Delta t}, \\
\Gamma_k &= \sum_{i \in S}\Gamma_{ik},} 
\end{equation*}
where $E_i$, $\omega_i$, $\vec{v_i}$, and $m_i$ are kinetic energy, sample weight (see Section~\ref{subsec:beam_selection}), velocity, and mass of the $i^{th}$ ion, respectively, and $S$ is a set of particles colliding with the $k^{th}$ mesh element.
$\Delta t$ is time duration of the injection of the NB source.
A total of 2.4 million NB ion samples are used, and any heat flux of the mesh elements with fewer than 3 sample ion collisions is excluded from the final accumulation results. 
In practice, the neglected portion of the total input beam power is $<$ 0.019\%.

To assess the maximum heat flux and the percentage of power deposition relative to the total input beam power for each PFC region (see Section~\ref{subsec:parameterScan}), we categorize and aggregate the heat flux data by the regions.
When defining the PFC regions, we utilize the assembly model (AM) \cite{schoonmaker2002cad} as a way of representing CAD data.
Specifically, we employ the AM system that composes the entire model with elements denoted as Parts and Assemblies.

In the AM, Parts are defined as pieces of the entire CAD model that divide it into mutually exclusive and exhaustively collective sections. 
Assemblies are containers holding these Parts or possibly lower-level Assemblies.
There is a Root Assembly that encompasses the entire model, and from this Root Assembly, a tree structure is formed where higher-level Assemblies recursively contain lower-level elements, establishing the classification system of the AM.

\begin{figure}[H]\centering
\includegraphics[width=0.9\textwidth]{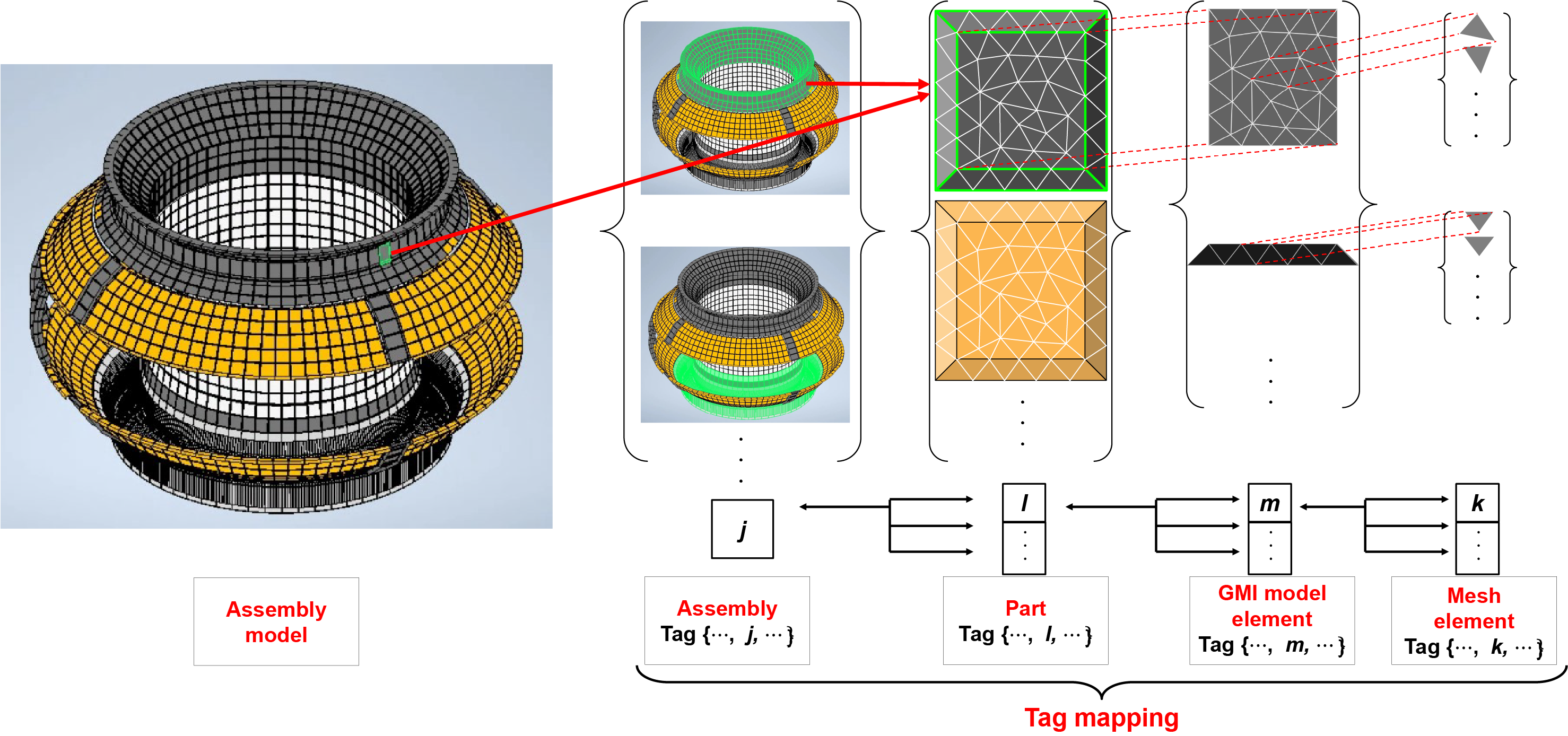}
        \captionsetup{width=\linewidth, format=hang}
        \caption{
                Schematic diagram of the hierarchical inclusion relationship among the AM, Assemblies/Parts, GMI model elements, and mesh elements with a detailed mapping of a Part to mesh elements via GMI model elements.
                }
        \label{fig:tagMapping}
\end{figure}

As illustrated in Figure~\ref{fig:tagMapping}, the tree structure enables the selection of the Assembly defining the target PFC region, subsequently facilitating the identification of its constituent Parts. Utilizing Simmetrix Simmodsuite's application programming interfaces (APIs) \cite{Simmetrix2024}, we can extract a comprehensive list of these targeted Parts along with their associated geometric model interface (GMI) elements and corresponding tags.  
Each GMI element is tessellated into mesh elements, and the list and tags of all these mesh elements can be acquired through the APIs of the Scientific Computation Research Center (SCOREC) core library \cite{ibanez2016pumi}. 
Utilizing these two mesh libraries, we obtain the tags of all mesh elements tessellating the targeted PFC region by using the tag of the corresponding Assembly \cite{moon2023component}.

\subsection{Simulation setup}
\begin{figure}[H]\centering
\includegraphics[width=0.7\textwidth]{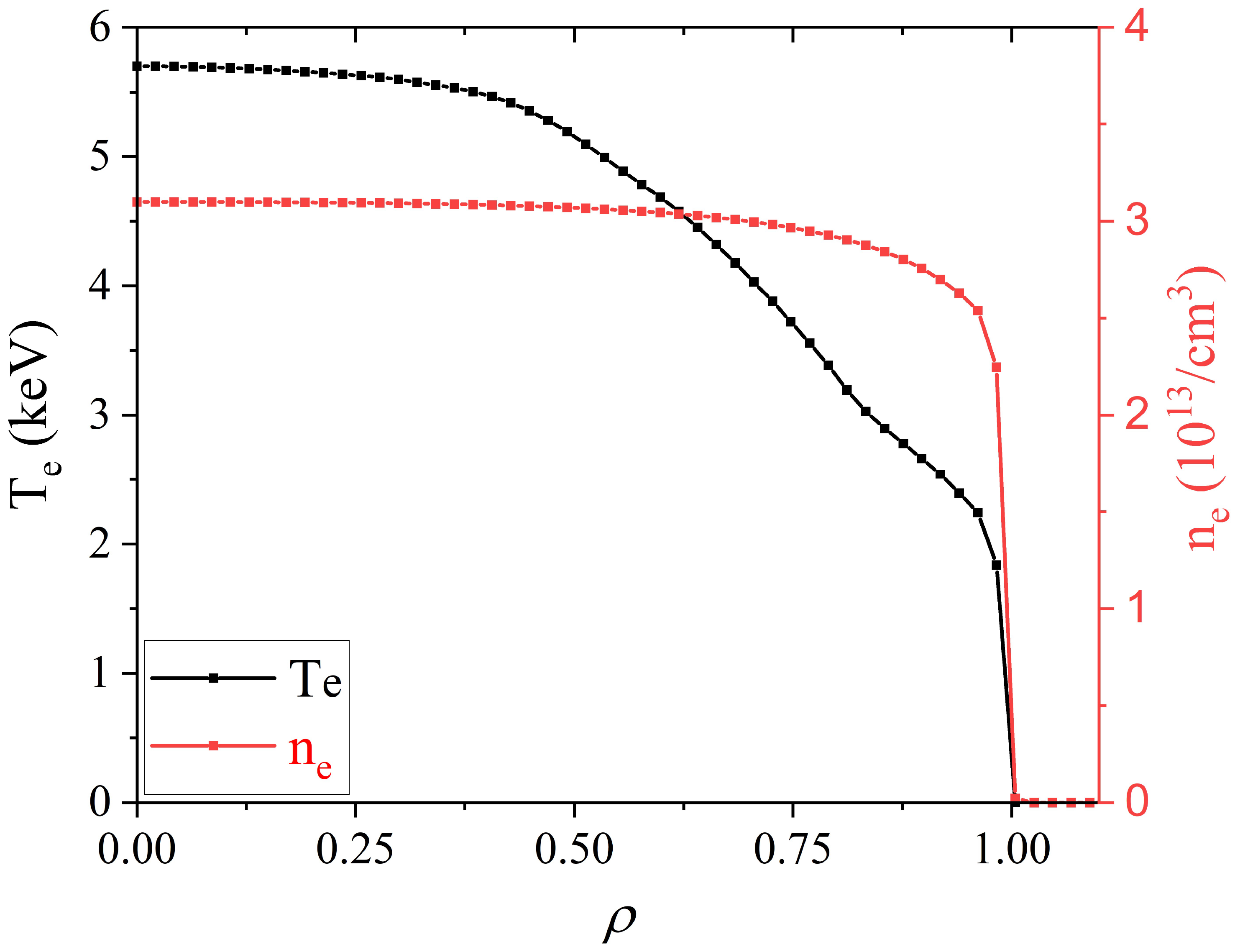}
        \captionsetup{width=\linewidth, format=hang}
        \caption{
                One-dimensional profiles of electron temperature ($\mathrm{T_e}$) and density ($\mathrm{N_e}$) over normalized poloidal flux.
                }
        \label{fig:electronDensityTemperatureProfile3D}
\end{figure}

In this study, we select sample beam particles based on the critical radial position of $\psi_N = 1.0$ following the way we described in Section~\ref{subsec:beam_selection}.
For beam deposition calculation, we set the electron density and temperature profiles as shown in Figure~\ref{fig:electronDensityTemperatureProfile3D} which are the typical H-mode profiles of KSTAR.
For each beam source, we utilize 2.4 million Monte Carlo particles to evaluate NB ion loss according to Section~\ref{sec:postProcesssing}.. 


KSTAR features two distinct NBI systems designated as NB1 and NB2. 
In this study, we focus on NB1 that has three ion beam sources, NB1-A, -B, and -C, directed through the pivotal position at the L port.
Each beam line of NB1-A, -B, and -C has a tangential radius of 1.486, 1.72, and 1.245~m, respectively \cite{Rhee_2019, oh2018progress}.
To obtain the general trend of the simulation results, the beam energies are set to minimum and maximum energies of 60~keV and 100~keV for the three different beam sources.
During sampling of neutral beam particles, subharmonic energies are not considered.

With the NBI setup, we investigate effects of $I_p$ and $\beta_p$ on NB ion loss. 
The $I_p$ is set to 0.50~MA, 0.75~MA, and 1.00~MA, and two cases of $\beta_p$ are examined, $\beta_p=1.0$ and $\beta_p=2.0$, giving a total of six equilibria tested for each beam source.
In these magnetic equilibria, the separatrix strike point is on the outer divertor, as illustrated in Figure~\ref{fig:magneticFieldfigure}. 
Another magnetic equilibrium with a different magnetic field structure has its strike point on the central divertor with $I_p = 1.00$ MA and $\beta_p = 1.0$. 
The parameters for these setups are detailed in Table~\ref{table:magneticFieldtable}.
Ripple is not considered due to the near-perfect axisymmetry of the KSTAR plasma
(magnetic ripple of $0.04\%$) \cite{park2018overview}.

\begin{figure}[H]\centering
\includegraphics[width=0.9\textwidth]{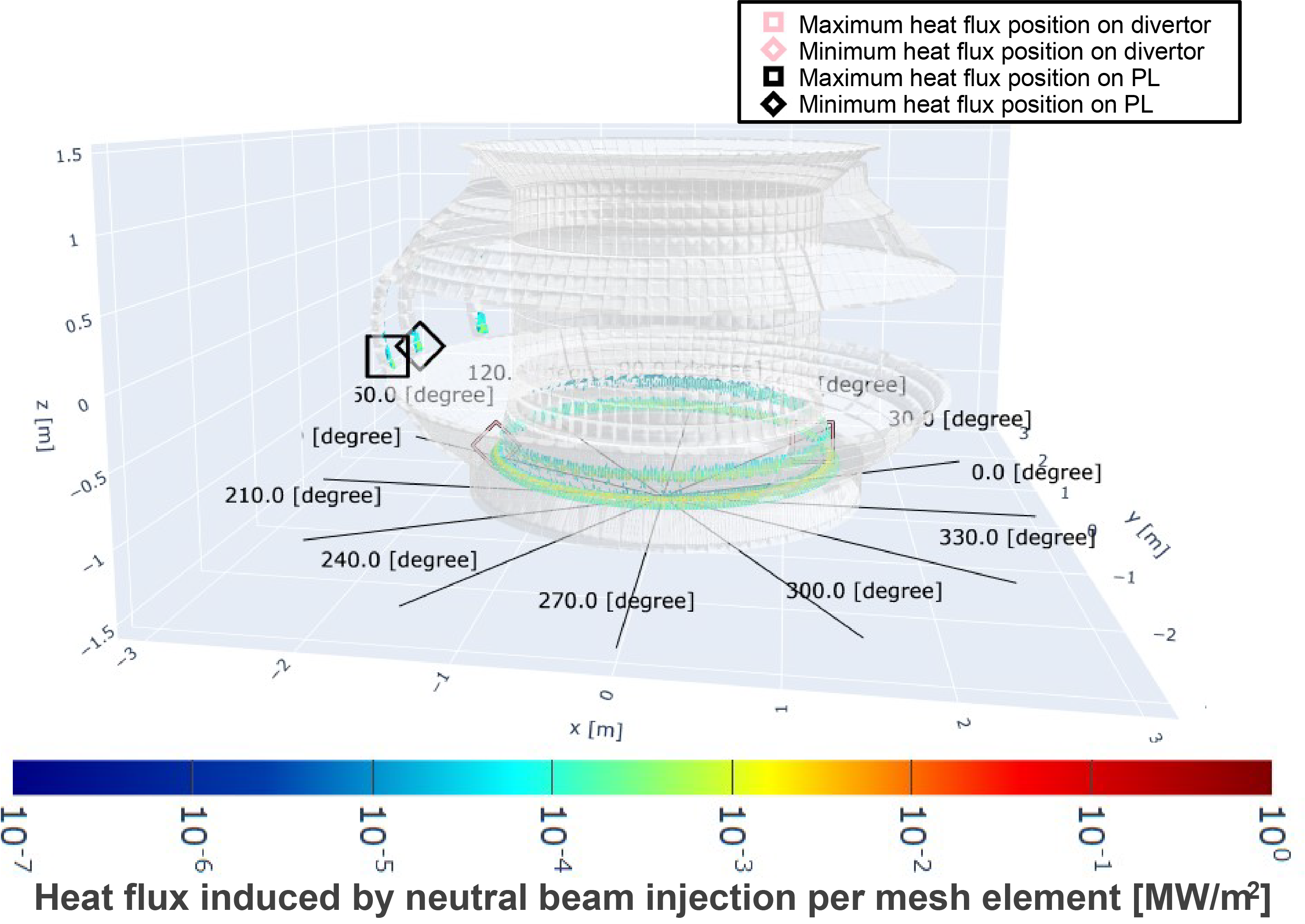}
        \captionsetup{width=\linewidth, format=hang}
        \caption{
                Three-dimensional visualization of heat flux due to NB ion loss for the reference case (Figure~\ref{fig:ReferenceCaseHowToUnfold}).
                }
        \label{fig:HF3Dview}
\end{figure}

\begin{figure}[H]\centering
\includegraphics[width=0.9\textwidth]{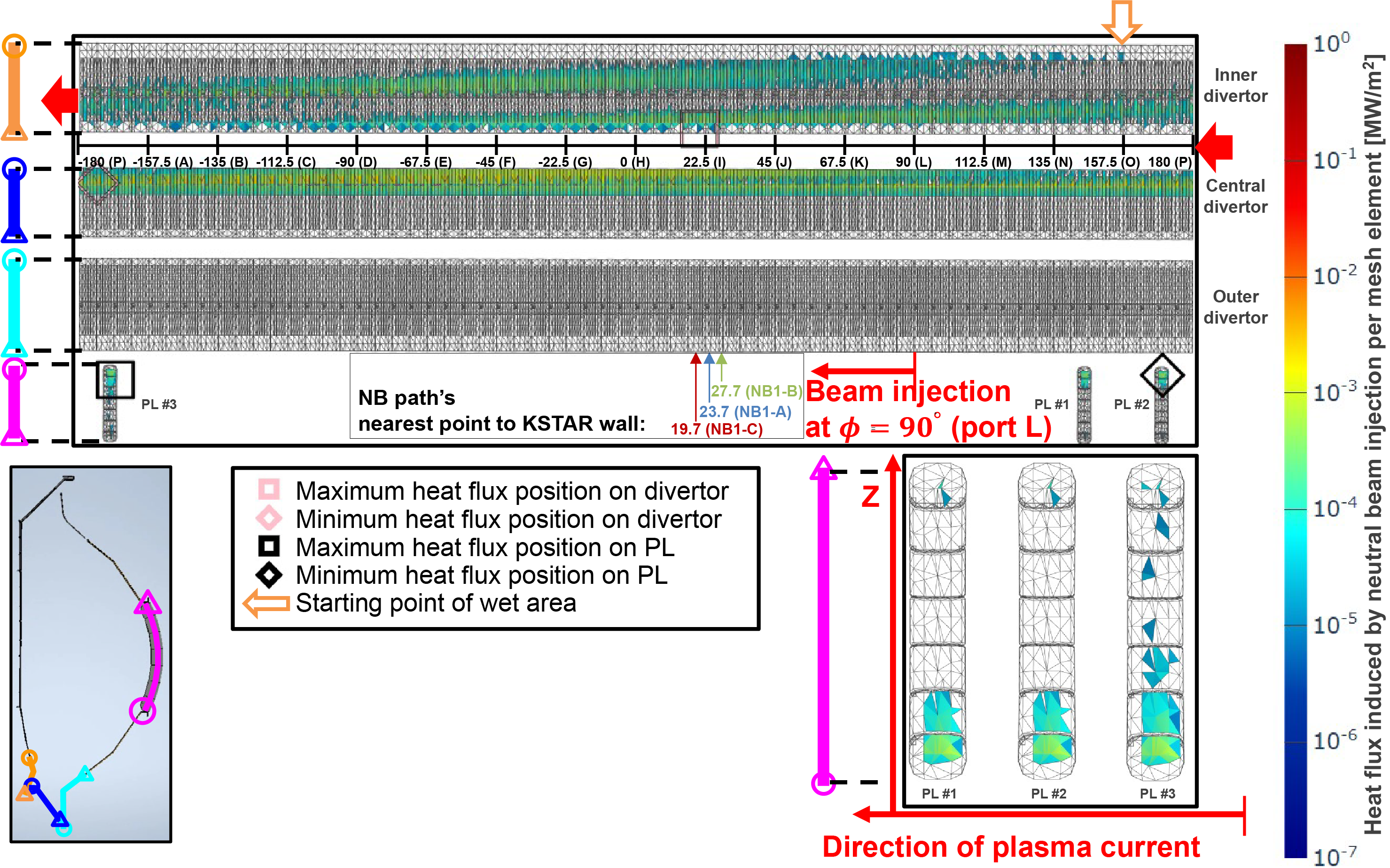}
        \captionsetup{width=\linewidth, format=hang}
        \caption{
                Heat flux colormap of the unfolded divertor and PL surfaces from the NuBDeC simulation results of the reference setup: 100~keV NB1-C neutral beam for the equilibrium case (I) in Figure~\ref{fig:magneticFieldfigure}.
                }
        \label{fig:ReferenceCaseHowToUnfold}
\end{figure}

\begin{figure}[H]\centering
\includegraphics[width=0.9\textwidth]{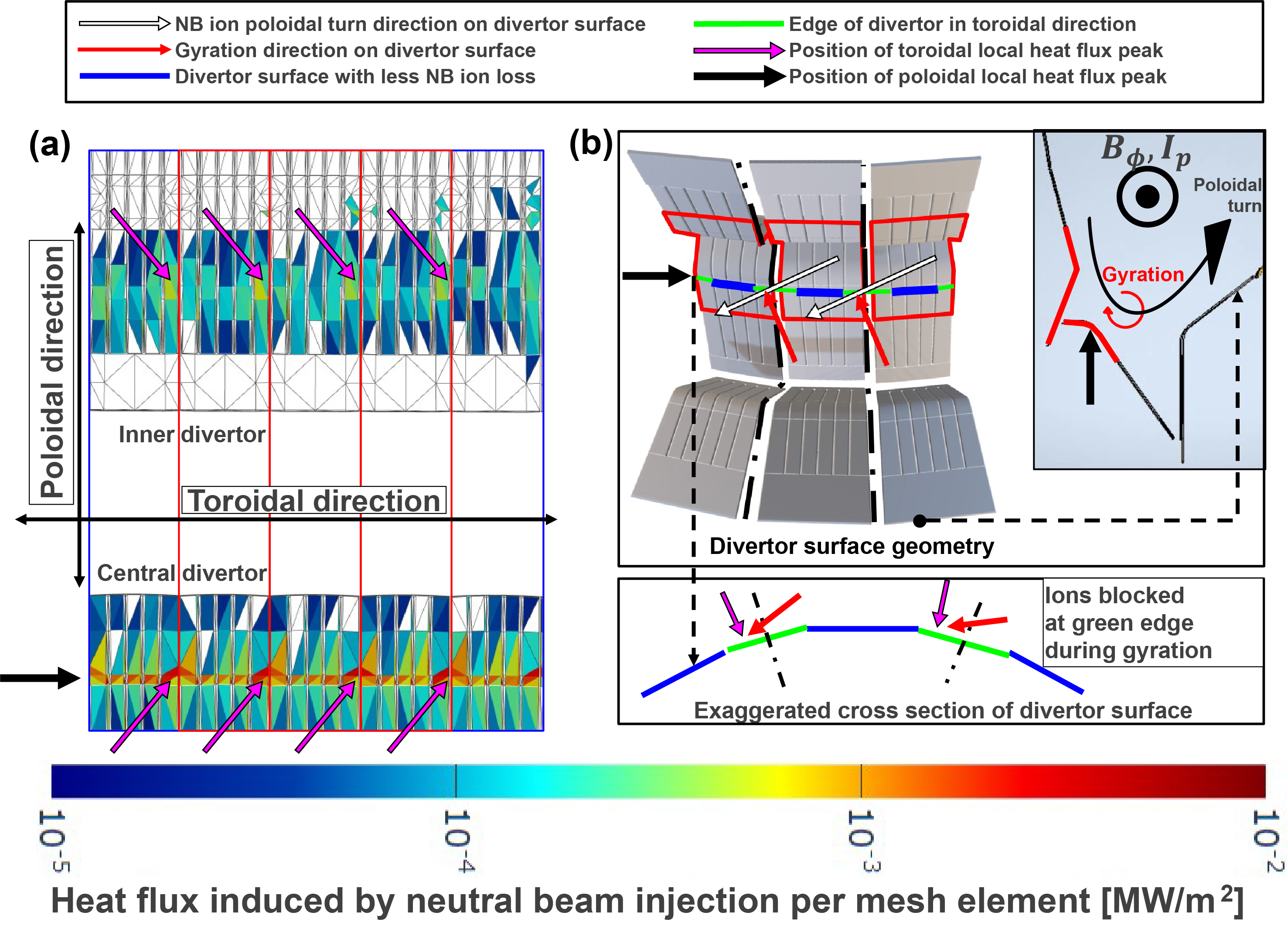}
        \captionsetup{width=\linewidth, format=hang}
        \caption{
                (a) Zoomed-in view of the inner and central divertor section between 130 and 160 degrees of the toroidal position in Figure~\ref{fig:ReferenceCaseHowToUnfold}.
                (b) The 3D geometry of the inner, central, and outer divertor surfaces. The three areas outlined by red boundaries on the surface correspond to the regions enclosed by the three red rectangles in the center of (a).
                }
        \label{fig:poloidalToroidalLocalHeatFluxExtremaWhy}
\end{figure}

\section{Results}~\label{sec:Results}
\subsection{Non-axisymmetric heat flux distribution}
The qualitative characteristics of the heat flux pattern obtained from the simulation results are investigated through a case study.
The reference setup is as follows: a 100~keV NB1-C neutral beam is sampled with the magnetic equilibrium formed at $I_p$ = 1.00~MA and $\beta_p$ = 1.0 and the outer strike point of the magnetic separatrix positioned at the outer divertor as shown by the thick red dashed line labeled (I) in Figure~\ref{fig:magneticFieldfigure}.
A full 3D view of the PFC surface heat flux observed is depicted in Figure~\ref{fig:HF3Dview},
with the wetted area colored according to heat flux levels.
The results indicate that all sections of the divertor (inner, central, and outer) are exposed to heat flux due to NB ion loss.

In Figure~\ref{fig:ReferenceCaseHowToUnfold}, detailed wetted areas are illustrated in an unfolded view of the divertor and PL. 
We unfold the toroidal angle (counterclockwise from the top view) horizontally and the poloidal angle (from the LFS to the HFS through the lower divertor) vertically. 
The vertical axis of the unfolded map points in the reversed direction of the poloidal angle, which increases counterclockwise.
In Figure~\ref{fig:ReferenceCaseHowToUnfold}, inner, central, outer divertors, and the PL are highlighted with orange, blue, cyan, and magenta bold lines on the PFC toroidal cross section, respectively, and illustrated in the bottom left panel. 
The directions of the PFC surface sections are indicated from circles to triangles in the corresponding colors. 
The PL is zoomed-in in the bottom right panel with the z-axis and plasma current directions marked with red arrows. 
The tangential points for the NB sources are marked in green (NB1-B), blue (NB1-A), and red (NB1-C). 
The section between 130 and 160 degrees of the toroidal position on the inner and central divertor surfaces is magnified in Figure~\ref{fig:poloidalToroidalLocalHeatFluxExtremaWhy}(a), with the area bordered by red lines featured on the divertor surface in Figure~\ref{fig:poloidalToroidalLocalHeatFluxExtremaWhy}(b).
A white arrow with an orange boundary points to the most upstream region in the NB ion trajectories or starting point of the plasma-wetted area, which extends diagonally from the upper right to the red arrow on the left side (the lower left of the inner divertor area in Figure~\ref{fig:ReferenceCaseHowToUnfold}). 
It continues from the red arrow on the right side (the lower right of the inner divertor area and the upper right of the central divertor area) spreading to the left.
The toroidal angle position is indicated between the inner and central divertor area.

The band-shaped wetted area extends obliquely because NB ions move both poloidally and toroidally as they drift inside the tokamak.
NB ions deposited (or ionized) deeper inside the tokamak, meaning at a smaller minor radius closer to the magnetic axis, require a longer path to collide with PFC surface because they need to move further for more drift toward the wall.
Then the wetted band expands diagonally over the PFC surfaces.

In Figure~\ref{fig:ReferenceCaseHowToUnfold}, the loss heat flux fades out towards the upper boundary of the band-shaped plasma-wetted area. This phenomenon occurs because, on the surface of the lower half PFC, particles ionized closer to the wall on the HFS are more likely to be lost on the surface near the HFS \cite{Rhee_2019}. As the number of ionized particles decreases closer to the HFS wall, the heat flux due to the loss on the PFC surface near the HFS also decreases. This variation in ionization distribution is related to changes in the beam flux density and ionization cross section distributions, which depend on the beam configuration and plasma profile. 

The extent of neutral beam ionization is governed by several factors. For the NB1-A source, the increased distance between its beam axis and the HFS wall results in reduced beam flux density due to beam divergence, consequently lowering the ionization rate. Conversely, while the NB1-C source's beam axis passes in close proximity to the HFS wall, ionization remains low due to the low plasma profile in this region—a condition that also affects NB1-A. As illustrated in Figure \ref{fig:electronDensityTemperatureProfile3D}, electron density diminishes towards the HFS wall, further suppressing the ionization rate \cite{suzuki1998attenuation}.
Further analysis of the band-like shape and heat flux distribution of the wetted area will be left for future work as it requires a deeper investigation into the relationship between ionization and loss distributions.

Additionally, local heat flux peaks occur along both poloidal and toroidal directions on the divertor as shown in the zoomed-in unfolded view in Figure~\ref{fig:poloidalToroidalLocalHeatFluxExtremaWhy}.
In Figure~\ref{fig:poloidalToroidalLocalHeatFluxExtremaWhy}(a), the magenta arrows point to local heat peaks in the toroidal direction, and the thick black arrow indicates a local heat peak at approximately 10~kW/m$^2$ in the poloidal direction.
In Figure~\ref{fig:poloidalToroidalLocalHeatFluxExtremaWhy}(b), the white arrows show the direction of ions' guiding-center orbits on the divertor surface while red arrows indicate the direction of gyration when ions graze the surface. 
The poloidal peaks show one order higher heat flux than the neighboring surfaces.
In the inset of the upper image of Figure~\ref{fig:poloidalToroidalLocalHeatFluxExtremaWhy}(b), the toroidal cross section view shows the surface with poloidal peak protrudes near the poloidal turn path.
The lower image in Figure~\ref{fig:poloidalToroidalLocalHeatFluxExtremaWhy}(b) shows a cross section along alternating blue and green lines representing the edge and middle surfaces of tungsten divertor modules, respectively. 
The blue middle surface is exposed to lower heat flux. 
The blue-green alternating line is divided with dash-dotted lines. 
Each section is the surface of an individual divertor module.

The toroidal heat flux distribution reveals a significant concentration at the leading edges of the tungsten divertor modules. This phenomenon arises from the divertor's non-axisymmetric structure, comprising 64 distinct modules \cite{kwon2021cfd}. When intersected by a horizontal plane coinciding with the alternating green-blue line in Figure~\ref{fig:poloidalToroidalLocalHeatFluxExtremaWhy}(b), the assembled divertor surface forms a 128-sided polygon rather than a circular cross-section. The lower portion of this figure provides an enlarged view of the polygon, with dash-dotted lines delineating individual modules.
This polygonal configuration results in protruding vertices, with each module featuring two such protrusions. The upstream protrusion relative to the neutral beam ion trajectories, where heat flux concentrates, is designated as the leading edge. This geometric feature, inherent to the modular design, significantly influences the localized heat flux distribution observed in the toroidal scan and results in toroidally repeated pattern of heat flux distribution and its local peak locations. 

The leading edges (marked with a magenta arrow in Figure~\ref{fig:poloidalToroidalLocalHeatFluxExtremaWhy}(b)) at the divertor module's edge of the divertor upstream in the toroidal direction prevents ions from reaching the downstream modules.
The surface of the downstream divertor module displays $\sim$10 times lower flux, as it only receives ions that were not lost at the leading surface. 
The heat flux concentration position varies depending on the geometry surrounding ion trajectory such as these proturdings.
This heat flux pattern aligns with the trends noted in \cite{ward2022locust}.

Regarding the PLs, relatively higher values of ion losses are observed at PL \#3, as shown in the bottom right panel in Figure~\ref{fig:ReferenceCaseHowToUnfold}. 
This occurs because PL \#3 is the most upstream PL in the NB ion trajectories, which the majority of NB ions first encounter. 
Consequently, only those NB ions survive at PL \#3 to continue to more downstream and can be lost at subsequent PLs.
Additionally, a pattern of reduced losses is particularly noticeable at PLs \#1 and \#2. 
Similar results were observed in \cite{Rhee_2019, oh2016overview}.

\subsection{Parameter scan} \label{subsec:parameterScan}
\begin{figure}[H]\centering
\includegraphics[width=0.9\textwidth]{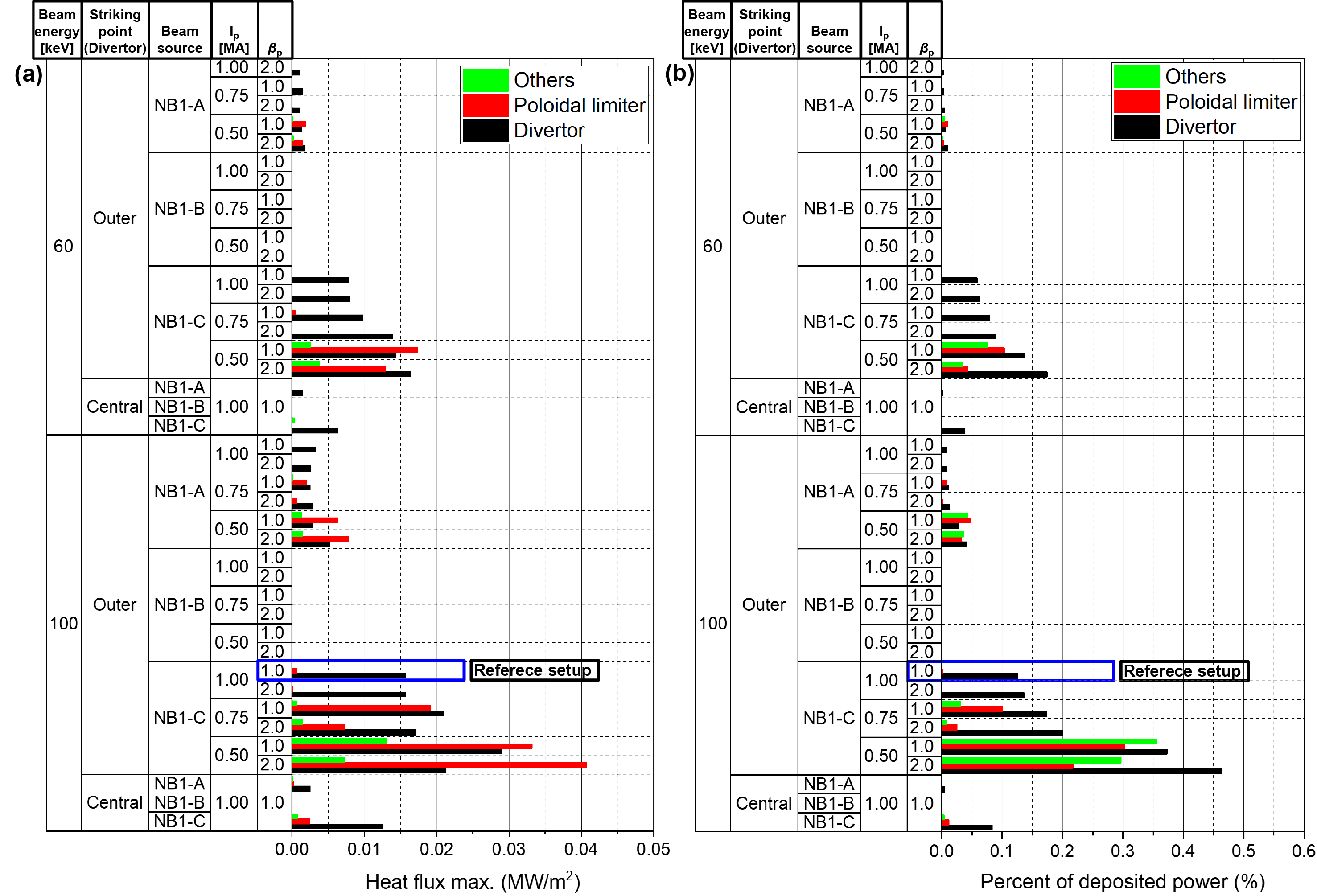}
        \captionsetup{width=\linewidth, format=hang}
        \caption{
                Maximum heat flux and percentage of power deposition for varying beam energies, beam sources, $I_p$, $\beta_p$, and outer strike point position of the magnetic separatrix on the divertor. The three PFC regions in the legend are illustrated in Figure~\ref{fig:poloidalCrossSectionView}.
                }
        \label{fig:HFSummary}
\end{figure}

The results of the parameter scan are summarized in Figure~\ref{fig:HFSummary}. 
It is observed that lower $I_p$ and higher $\beta_p$ values lead to an increased percentage of power deposition. 
Furthermore, the maximum heat flux and the percentage of power deposition vary with both the beam energy and beam source. 
Specifically, these quantities increase in the order NB1-B $<$ NB1-A $<$ NB1-C and also increase for higher beam energy.

\begin{figure}[H]\centering
\includegraphics[width=0.8\textwidth]{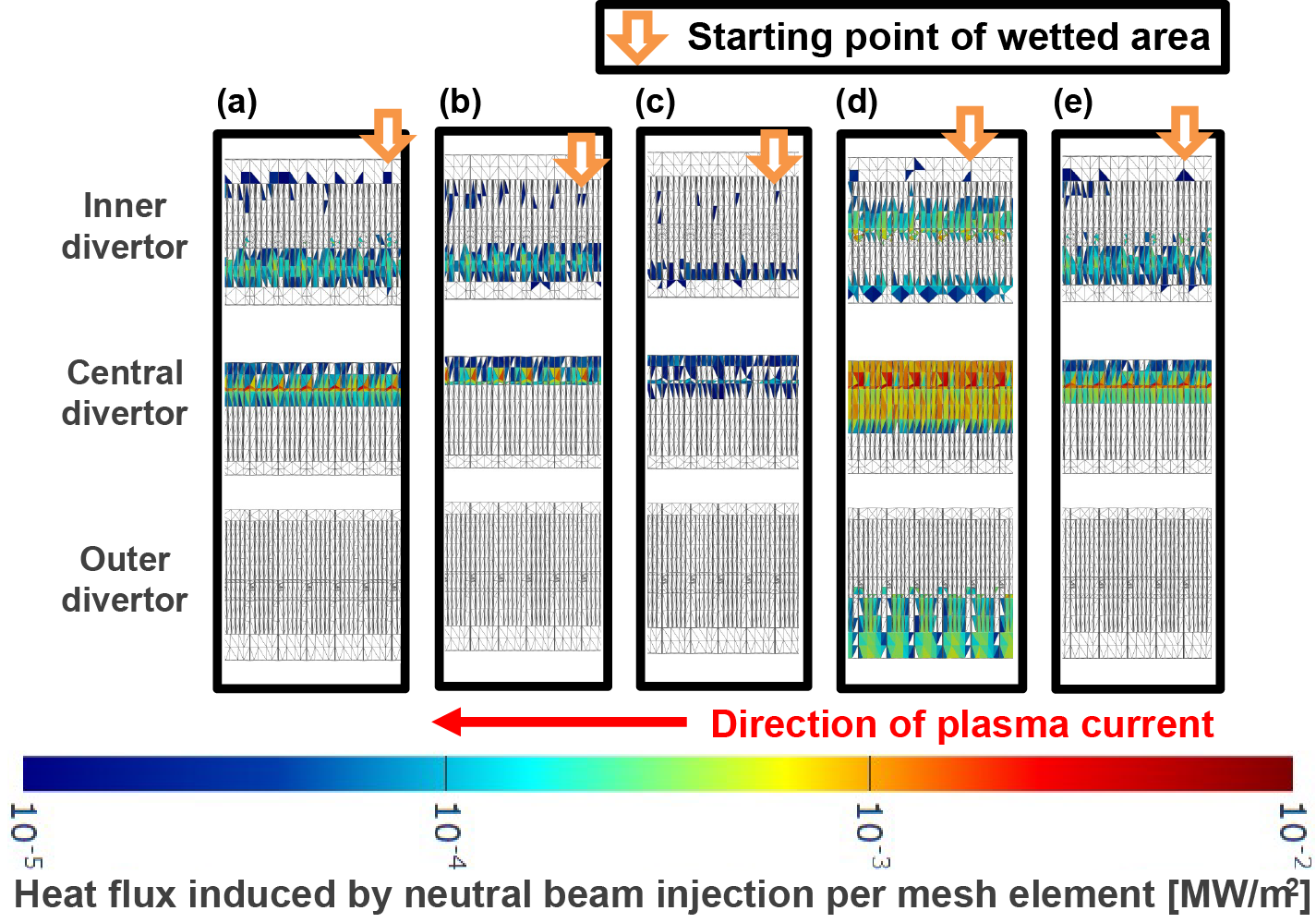}
        \captionsetup{width=\linewidth, format=hang}
        \caption{
                Enlarged planar views of the unfolded divertor comparing NuBDeC heat flux distributions due to NB ion loss: (a) reference setup, and single-parameter variations of (b) 60~keV beam energy, (c) NB1-A beam source, (d) $I_p$ = 0.50~MA, and (e) $\beta_p$ = 2.0. A white arrow with an orange boundary points to the most upstream point in the NB ion trajectories or starting point of the plasma-wetted area.
                }
        \label{fig:RefVariationUnfold}
\end{figure}

When the beam energy is reduced from 100~keV to 60~keV, the maximum heat flux decreases by approximately 75--80\% and the percentage of power deposition decreases by about 90--95\% as depicted in Figure~\ref{fig:HFSummary}. 
Furthermore, a narrower poloidal wetted area forms on the divertor surface as illustrated in Figure~\ref{fig:RefVariationUnfold}(b).
Decreasing the beam energy results in smaller curvature and $\nabla B$ drifts, which makes the poloidal turn path less likely to reach the PFCs from inside of the LCFS.

\begin{figure}[H]\centering
\includegraphics[width=0.8\textwidth]{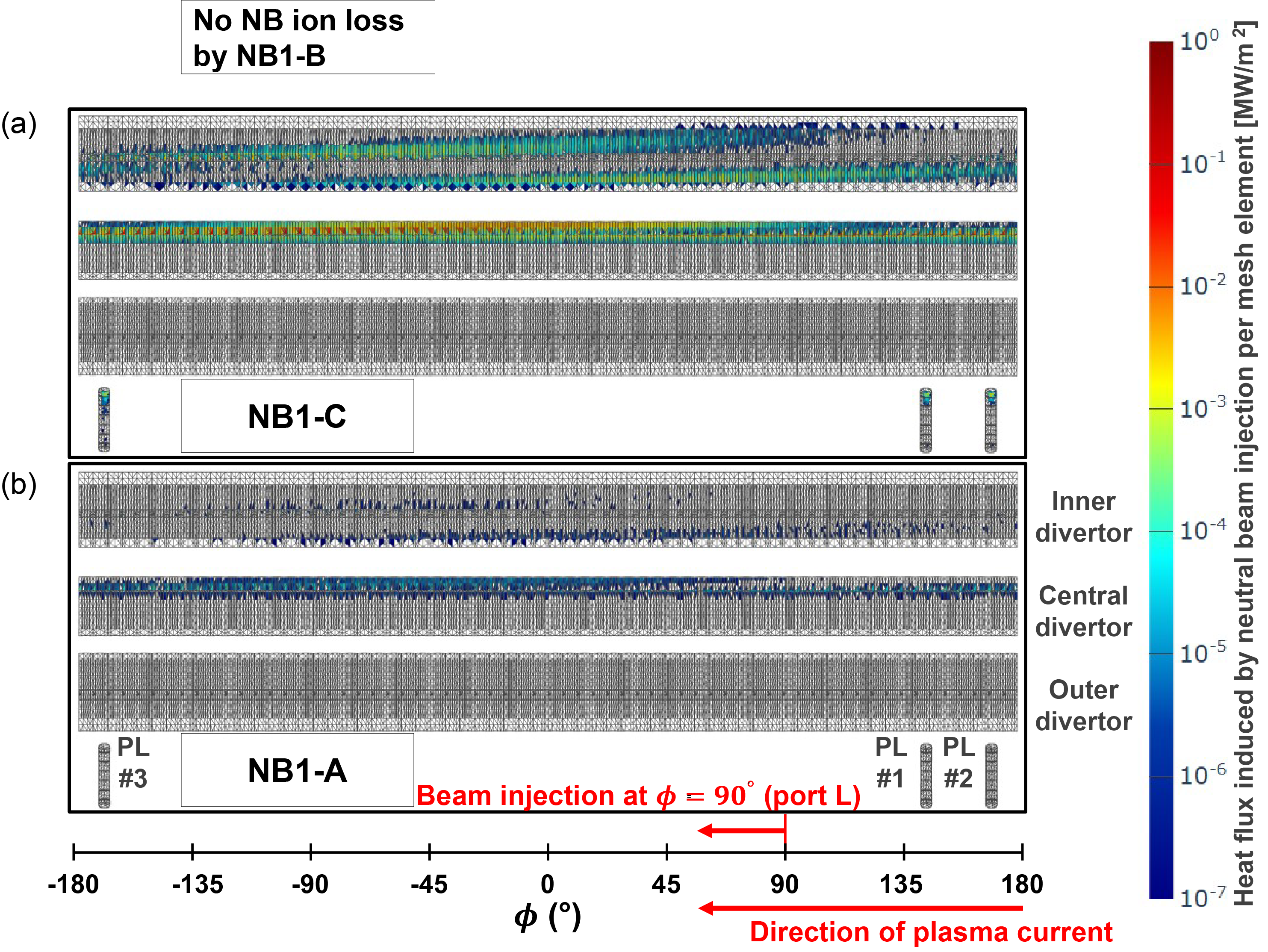}
        \captionsetup{width=\linewidth, format=hang}
        \caption{
                 Variations in beam sources from NuBDeC simulation parameter scans showing (a) the reference setup and (b) a single-parameter variation to the NB1-A beam source. There is no identified NB ion loss by NB1-B.
                }
        \label{fig:energyVariationNB1AC}
\end{figure}

\begin{figure}[H]\centering
\includegraphics[width=0.8\textwidth]{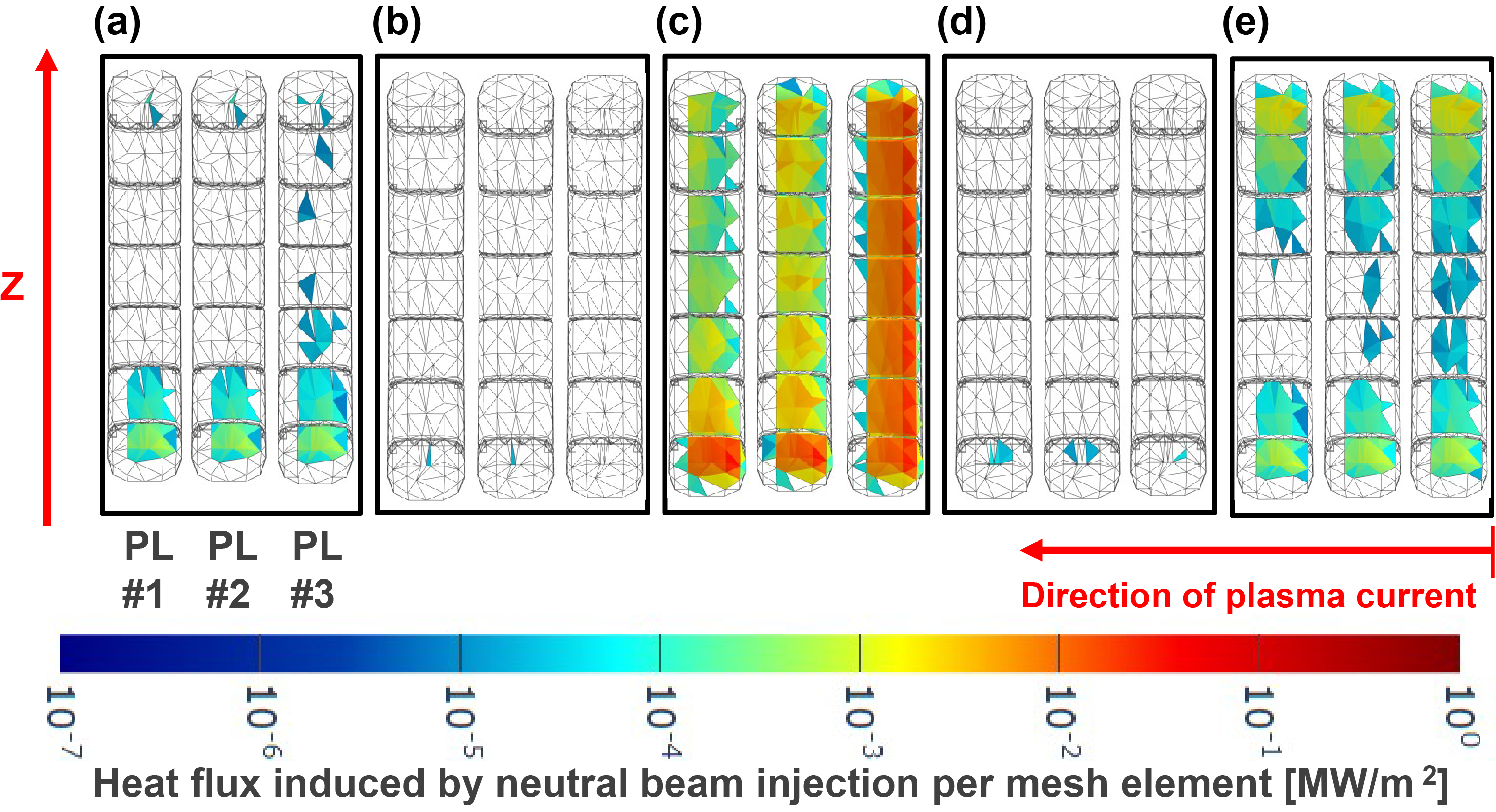}
        \captionsetup{width=\linewidth, format=hang}
        \caption{
                Enlarged planar views of the unfolded PL comparing NuBDeC heat flux distributions due to NB ion loss: (a) reference setup, and single-parameter variations of (b) NB1-A beam source, (c) $I_p$ = 0.50~MA, (d) $\beta_p$ = 2.0, and (e) outer magnetic separatrix strike point on the central divertor. Heat flux on the PL disappears when the beam source is changed from the reference case to a 60~keV NB1-C beam.
                }
        \label{fig:RefVariationUnfoldPL}
\end{figure}

Changing the NB source in the reference case shows a decrease in both maximum heat flux and power deposition percentage as the tangential radius increases (from NB1-C to NB1-A) as shown in Figure~\ref{fig:HFSummary}. 
The deepest beam source NB1-B does not result in any NB ion losses on the PFCs. 
In contrast, NB1-A and NB1-C show NB ion losses, where the deeper deposition source NB1-A generates three to four times less NB ion losses compared to the shallowest NB1-C source.
The NB1-A beam source results in less ionizatoin occurring toward the HFS. 
Consequently, ion losses tend to occur more downstream of the poloidal turn, as shown in Figures~\ref{fig:energyVariationNB1AC}(b), \ref{fig:RefVariationUnfold}(c), and \ref{fig:RefVariationUnfoldPL}(b).
The NB1-B case is excluded from further consideration because no NB ion loss was found in the simulation results.
The dependency of NB ion loss on the beam energy and beam source is consistent with the results of \cite{Rhee_2019}.

For all simulation parameters including $\beta_p$ values, beam sources, and beam energies reducing the $I_p$ from 1.00~MA to 0.50~MA produces more than twice the maximum heat flux and percentage of power deposition on the divertor and PL regions. 
At lower $I_p$s, the wetted area on the divertor surface is more widely distributed toward the ion's poloidal turn direction compared to the reference case. 
Also, an additional wetted area appears on the outer divertor, as shown in Figures~\ref{fig:RefVariationUnfold}(d) and \ref{fig:RefVariationUnfoldPL}(c).

As $I_p$ diminishes, the poloidal magnetic field weakens, resulting in an elevated safety factor. 
An increase in the safety factor means that the guiding center of the particles takes more toroidal turns per poloidal turn, leading to more pronounced drift toward the wall.
This change in safety factor allows the poloidal turn orbit to be close to the central divertor and the inner divertor, leading to the increased ion losses shown in Figures~\ref{fig:RefVariationUnfold}(d) and \ref{fig:RefVariationUnfoldPL}(c).
This causes an extension of the wetted area along the poloidal direction on the central divertor.

Increasing $\beta_p$ from 1.0 to 2.0 while maintaining $I_p$ at 1.00~MA results in an approximately 10\% increase in the percentage of power deposition on the divertor though the heat flux on the PL vanishes. 
For $I_p$ values of 0.50~MA and 0.75~MA with the increase in $\beta_p$ from 1.0 to 2.0, the percentage of power deposition on the PL and other regions decreases by 10\%. 
Although the maximum heat flux trend varies by region at these lower $I_p$ values, the differences remain within 10\%. As $\beta_p$ varies the Shafranov shift changes the structure of the magnetic field. 
In other words, the magnetic axis moves away from the center of tokamak, which leads to increased ion losses as shown in Figure~\ref{fig:HFSummary}.

\begin{figure}[H]\centering
\includegraphics[width=0.8\textwidth]{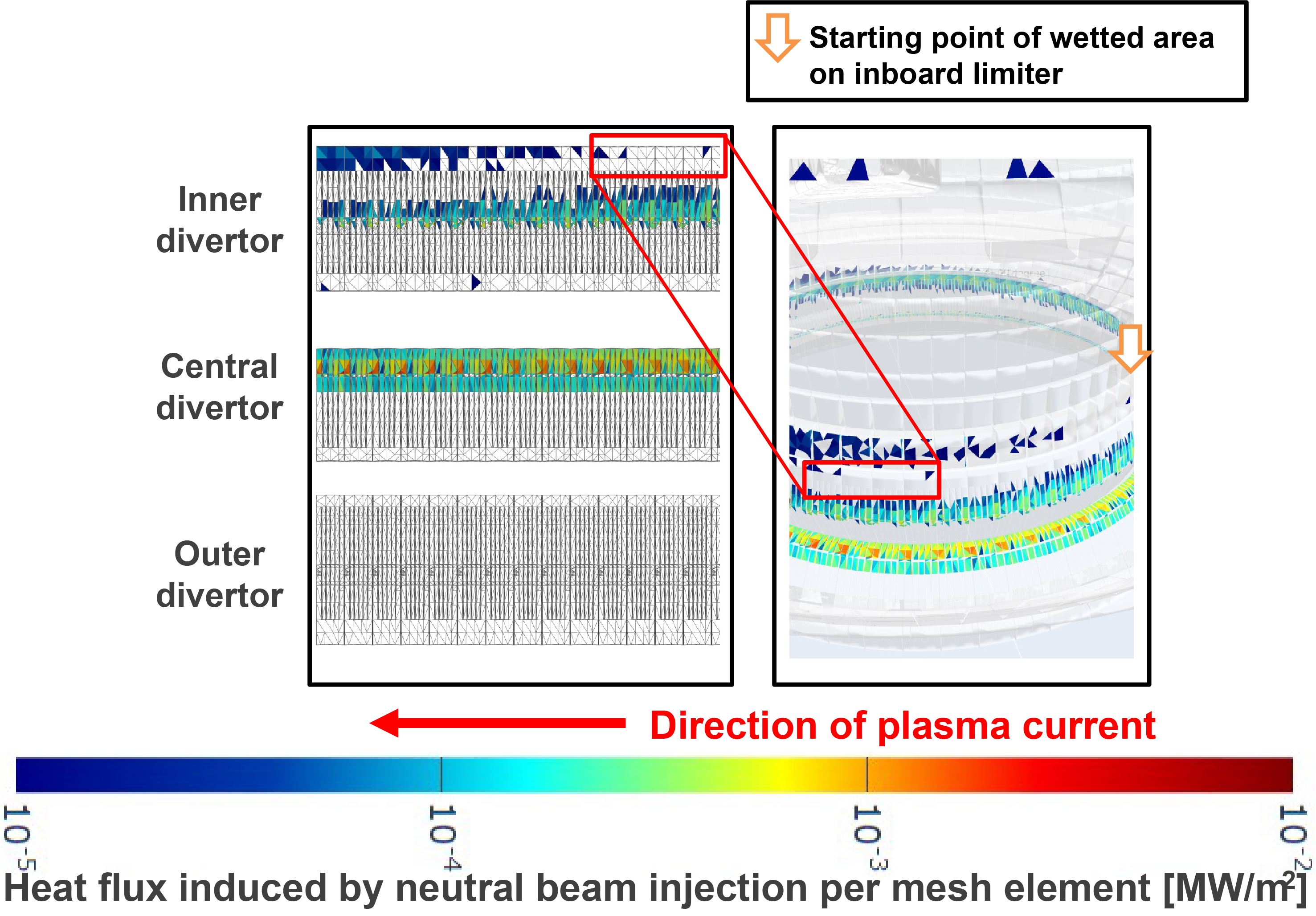}
        \captionsetup{width=\linewidth, format=hang}
        \caption{
                Enlarged planar view of the unfolded divertor with the outer strike point of the separatrix positioned on the central divertor, compared to NuBDeC NB ion loss heat flux distributions in Figure~\ref{fig:RefVariationUnfold}. Note that the wetted area reaches the inner limiter. A white arrow with an orange boundary points to the most upstream point in the NB ion trajectories or starting point of the plasma-wetted area.
                }
        \label{fig:RefVariationUnfoldStrikingPoint}
\end{figure}

When the magnetic separatrix's outer strike point moves from the outer divertor to the central divertor, higher maximum heat flux and percentage of power deposition are observed on the overall divertor while lower values are confirmed on the PL as shown in Figure~\ref{fig:HFSummary}. 
The wetted area on the divertor surface extends further in the direction of the strike point movement, eventually reaching the inboard limiter as shown in Figure~\ref{fig:RefVariationUnfoldStrikingPoint}. 
A wider wetted area is observed on the PL with the distribution biased toward a higher Z position as observed in Figure~\ref{fig:RefVariationUnfoldPL}(e).

Moving the outer strike point of the magnetic separatrix from the outer to the central divertor causes a larger variation in the magnetic field at the plasma edge and the scrape-off layer (SOL) region than the variation caused by the Shafranov shift due to changes in $\beta_p$ (Figure~\ref{fig:magneticFieldfigure}). 
The NB ions that make poloidal turns toward the plasma edge and in the SOL region are more likely to be lost.
Consequently, the change in magnetic field structure by varying the strike point position has more impact on NB ion loss than that resulting from changes in $\beta_p$ (Figures~\ref{fig:RefVariationUnfoldPL}(e) and \ref{fig:RefVariationUnfoldStrikingPoint}).

In the parameter scan, remarkable tendencies and value ranges of the heat flux are as follows.
The most pronounced NB ion loss heat flux is generated by NB1-C, which impinges nearest to the inner wall on the HFS and has the shortest tangential radius.
The percentage of power deposition increases with lower $I_p$ and higher $\beta_p$ values of magnetic equilibrium conditions.
The range of the maximum heat flux values and the percentage of power deposition for each PFC region (Figure~\ref{fig:poloidalCrossSectionView}) are summarized in Table~\ref{table:MaxHFPowerDepoRange}.

\begin{table}[bp]
\caption{Range of maximum heat flux and percentage of power deposition relative to the total input beam power (1 MW) for each PFC region out of 42 NuBDeC simulation setups using 60/100~keV NB1-A, -B, -C (1 MW of power) across seven magnetic equilibria. }
\label{table:MaxHFPowerDepoRange}
\begin{indented}
\lineup
\item[]\begin{tabular}{@{}*{4}{c}}
\br
                                    & Divertor     & PL & Others\\
\mr
Maximum heat flux~[kW/m$^2$]        & $<$ 30   & $<$ 40       & $<$ 15\\
Power deposition [\%] & $<$ 0.5   & $<$ 0.3       & $<$ 0.35\\
\br
\end{tabular}
\end{indented}
\end{table}

In our case study, we focus on the main parameters that control the heat flux patterns, which are summarized as below.
\begin{enumerate}
\item The poloidal turn orbit and correspondingly the poloidal loss position are presumed to be determined by the beam deposition (i.e., ionization) position. 
For instance, an ion ionized near the plasma edge proximal to the HFS wall traverses a poloidal turn orbit at a larger minor radius.
In this case, its trajectory collides eariler with the PFC surface after a shorter distance from its beam deposition position. 
In this study, we used beam sources with different tangential radii of beam injection. 
Notably, when the beam is injected closer to the HFS, the beam deposition also occurs closer to the HFS for the given plasma profiles.
Moreover, variations in $I_p$ and $\beta_p$ lead to changes in the magnetic flux surface, which in turn causes redistribution of the plasma profile and consequently a shift in the beam deposition position.
\item Shifts of the magnetic flux surfaces and the magnetic separatrix’s outer strike point tend to cause NB ion losses at different positions, even with identical beam deposition distribution.
\item Radial drift of NB ions causes a deviation ($\Delta \sim q v_{\parallel}$) of the center of the poloidal turn orbit \cite{Rhee_2019}. 
$v_{\parallel}$ tends to be higher with increased beam energy. The safety factor increases due to weaker poloidal magnetic field induced by the lower $I_p$. 
Therefore, higher beam energy and lower $I_p$ cause the particles to drift more, which results in higher heat flux on divertor.

\end{enumerate}

\section{Conclusion} \label{sec:Conclusion}
Results of the investigation into the characteristics of the heat flux pattern through a parameter scan provide significant insights into the influence of various parameters on NB ion losses and their resultant heat flux distributions. 
The reference case serves as a baseline for evaluating changes in heat flux distribution across varying parameters.
The analysis of unfolded PFC surfaces demonstrates non-axisymmetric heat flux patterns, particularly showing band-shaped wetted areas that obliquely extend along the direction of poloidal and toroidal turns of NB ions. 
These patterns are attributed to the beam deposition and subsequent drift along the magnetic field lines. 
The heat flux forms peaks due to the concentration of NB ion losses at protruding surfaces and toroidal leading edges.

The parameter scan reveals that lower $I_p$ and higher $\beta_p$ values result in increased power deposition on the divertor. 
Additionally, varying the beam source shows that shallower beam deposition (as with NB1-C) results in more NB ion losses compared to deeper sources like NB1-A and NB1-B.
Reducing the beam energy from 100~keV to 60~keV significantly decreases both the maximum heat flux and the percentage of power deposition, forming narrower poloidal wetted areas on the divertor surface.

Movement of the magnetic separatrix's outer strike point from the outer to the central divertor significantly changes the heat flux distribution. The wetted area extends further in the direction of the strike point movement and reaches the inboard limiter. 
This movement also results in a wider wetted area on the PL, biased toward higher Z positions. 
Changes in the strike point position induce larger variations in the magnetic field at the plasma edge and the SOL region compared to variations caused by changes in $\beta_p$.

Overall, these findings can support the optimization of magnetic confinement and beam injection strategies to minimize NB ion losses and manage heat flux on PFCs in conventional tokamaks including KSTAR, contributing to the advancement of fusion reactor design and operation.
In future work, impact of spatially varying distributions of the beam deposition on the NB ion loss can be explored. 
This research will enable us to determine which parameters need adjustment to shift and evenly spread the heat flux, preventing its localization.
For example, by back-tracing the path of lost NB ions, we can investigate the correlation between the distributions of the beam deposition and the NB ion loss in 3D space. 
Then understanding the correlation with the beam deposition distribution influenced by factors such as beam injection direction, beam energy, and ionization cross section distribution will allow us to identify and analyze factors that make specific shapes of heat flux patterns and unnecessarily concentrate heat flux on local areas with specific NBI operation scenario.

In addition, as a follow-up study, we can investigate underlying factors behind the influence of $\beta_p$ variation. 
Understanding the correlations quantitatively, particularly why changes in $\beta_p$ do not have a significant effect as $I_p$ changes or have different effects depending on the range of $I_p$, could help optimize NBI operations.
It was found that the percentage of power deposition on the divertor increases as the $I_p$ decreases and $\beta_p$ increases. 
However, for the PL, the percentage of power deposition can either increase or decrease with a decrease in $\beta_p$ as the $I_p$ value changes (Figures~\ref{fig:HFSummary} and \ref{fig:RefVariationUnfoldPL}(d)).
Unveiling the reasons behind this behavior and observing how the power deposition ratio varies across different PFCs could provide valuable insights. 

By using a tool for quantitative analysis in a virtual environment of locally concentrated heat flux distributions based on actual geometry and experimental measurements, detailed NB ion collision information over the KSTAR PFCs can be provided. It can help analyze the thermal, chemical, and mechanical degradation, as well as atomic processes over the PFCs by providing NB-ion loss particle and energy flux information. Ultimately, it may be possible to help optimizing H-mode operation scenarios for the KSTAR device with a tungsten divertor and support research and development of the advanced scenario for commercial fusion devices.

\ack

We thank Mark Shephard at Rensselaer Polytechnic Institute and Mark Beall at Simmetrix Inc. for their consistent support of mesh softwares, PUMI (Parallel Unstructured Mesh Infrastructure) and SimmodSuite. This research was supported by the R\&D Program through the Korea Institute of Fusion Energy (KFE) funded by government funds of the Republic of Korea (KFE-EN2341) and by a National Research Foundation of Korea (NRF) grant funded by the Korean government (MSIT) (No. RS-2023-00254695). Computing resources were provided from the KFE supercomputer, KAIROS, funded by the Ministry of Science and ICT of the Republic of Korea (MSIT) (KFE-EN2341-9).

\bibliography{manuscript_ESY20240817}

\end{document}